\documentclass[11pt]{article}
\usepackage[a4paper,margin=28mm]{geometry}
\usepackage{amsmath,amssymb,mathtools,bm}
\usepackage{mathrsfs}
\usepackage{microtype}
\usepackage{breqn}
\usepackage{amsthm}
\usepackage[hidelinks]{hyperref}
\usepackage[nameinlink,capitalize]{cleveref}
\usepackage[numbers,sort&compress]{natbib}
\usepackage{authblk}
\usepackage[T1]{fontenc}
\usepackage{comment}

\theoremstyle{plain}
\newtheorem{theorem}{Theorem}[section]
\newtheorem{proposition}[theorem]{Proposition}
\newtheorem{corollary}[theorem]{Corollary}

\theoremstyle{definition}
\newtheorem{definition}[theorem]{Definition}
\theoremstyle{remark}

\newcommand{\ii}{\mathrm i}
\newcommand{\ee}{\mathrm e}
\newcommand{\dd}{\mathrm d}
\newcommand{\Disc}{\operatorname{Disc}}

\newcommand{\AUF}{\mathrm{\scriptscriptstyle AUF}}

\newcommand{\Gstar}{\Gamma^{\!*}}
\newcommand{\RE}{\operatorname{Re}}
\newcommand{\IM}{\operatorname{Im}}
\newcommand{\Log}{\operatorname{Log}}
\newcommand{\Arg}{\operatorname{Arg}}

\makeatletter
\newcommand*{\defeq}{\mathrel{\rlap{%
                     \raisebox{0.3ex}{$\m@th\cdot$}}%
                     \raisebox{-0.3ex}{$\m@th\cdot$}}%
                     =}
\makeatother

\begin{document}

\title{Resurgence without the Borel Plane: Mellin--Barnes Summation and Analytic Stokes Reconstruction}
\author{Roberto Ricci\footnote{E-mail: roberto.ricci@enea.it (Roberto Ricci)}}
\affil[1]{ENEA, Nuclear Department NUC-DTT, Frascati Research Center, Via E. Fermi 45, 00044 Frascati (Rome), Italy}
\date{}
\setcounter{Maxaffil}{0}
\renewcommand\Affilfont{\itshape\small}
\maketitle

\begin{abstract}
We develop a route to \emph{resurgence without the Borel plane}:
sums of divergent perturbative series and their encoded nonperturbative monodromy are
constructed directly in complex perturbative order.  In the Analytic
Umbral Framework (AUF), an analytic coefficient clock produces the
sum through a Mellin--Barnes integral and its
monodromy through a bare inverse Mellin transform.  Neither a Borel
transform nor continuation of a Borel germ is required for this
construction.  For Watson-admissible Gevrey clocks, the completion
agrees with the canonical Borel--Laplace sum in the guaranteed
directions.  When the admissible aperture exceeds $2\pi$, the clockwise
difference of adjacent logarithmic determinations is given by
$2\pi\ii$ times the inverse-Mellin carrier.  This carrier is also
the AUF completion of a null clock and is flat on the overlap. The sum and its aggregate monodromy are thus obtained
``in one shot'', before resolving individual actions and
fluctuation series.

Euler's equation, zero-dimensional quartic models, the scaled Gamma function and a
solvable $O(4)$ toy model exemplify the procedure, and show how further analysis recovers
directional Stokes variations.  The lognormal benchmark extends
the construction beyond every finite Gevrey order: no finite-order
Borel transform gives a convergent germ, yet explicit clock-selection
conditions yield an AUF completion and a nonzero ordinary flat
carrier.  Distributed relaxation supplies a physical realization
of this growth class.  AUF therefore both reconstructs familiar
resurgent information without Borel-plane continuation and extends
completion beyond finite-order Borel--Laplace summation.
\end{abstract}

\noindent\textbf{Keywords:} resurgence; Mellin--Barnes integrals;
divergent perturbation theory; Analytic Umbral Framework;
Stokes phenomena; transseries; non-Gevrey series; lognormal series.

\section{Introduction}
\label{sec:introduction}

This paper develops a route to \emph{resurgence without the Borel
plane}.  An admissible analytic continuation in complex order of the 
coefficients of a divergent perturbative series supplies both a sectorial sum and
an exact carrier of its encoded nonperturbative monodromy via analytic umbral pairings, represented by Mellin--Barnes integrals.  We specify the analytic conditions for this construction and
demonstrate it explicitly in benchmarks of increasing complexity,
showing how directional Stokes variations can be recovered from
the resulting analytic functions.  Neither a Borel transform nor continuation
of a Borel germ is required for these constructions.  The sum and
its aggregate monodromy are obtained ``in one shot'', before
separating individual actions and fluctuation series.

The motivation comes from divergent perturbation theory in physics,
where large-order growth can encode nonperturbative information.
Classical examples include Dyson's argument for quantum
electrodynamics, Bender and Wu's analysis of the anharmonic
oscillator, and Lipatov's semiclassical analysis of quantum field
theory \cite{Dyson1952,BenderWu1973,Lipatov1977}.  Factorial growth
is both an obstacle to direct summation and a source of analytic
information \cite{Dingle1973,Costin2009}.

The standard resurgent route first Borel transforms the formal
series, continues the Borel germ, and reconstructs sectorial
functions by directional Laplace integration.  Borel singularities
encode actions and Stokes data, while alien calculus organizes
relations among asymptotic sectors
\cite{Ecalle1981,Costin2009,MitschiSauzin2016,Dorigoni2019,Aniceto2019}.
Applications span quantum mechanics, matrix models, string theory
and quantum field theory \cite{Marino2015,DunneUnsal2016,Aniceto2019}.
The Analytic Umbral Framework (AUF) \cite{RicciAUF}, an analytic extension of formal umbral methods
\cite{BabusciEtAl2019Methods,Ricci2026UmbralTheory}, places the construction of sums
and their monodromy in perturbative-order/Mellin space.  Under the
stated Gevrey hypotheses, Borel--Laplace theory identifies the
resulting function with the canonical sum; the non-Gevrey benchmark
shows that the AUF construction also applies beyond that comparison.

Mellin--Barnes methods already provide representations and
asymptotic expansions of special functions and Feynman amplitudes
\cite{ParisKaminski2001,FriotGreynatDeRafael2005}.  Their use in
hyperasymptotic analysis \cite{FriotGreynat2010}, and the recovery
of analytic information through hypergeometric and Meijer--$G$
approximants \cite{MeraPedersenNikolic2018}, form part of this
background.  The contribution here is the systematic use of an
analytic coefficient clock to construct both the sum and its
monodromy, with explicit admissibility and flatness conditions.

Exact Perturbation Theory offers a complementary comparison:
for classes of quantum-mechanical systems, a reorganization of
perturbation theory permits Borel summation to the full result
without introducing separate transseries sectors
\cite{SeroneSpadaVilladoroInstantons2017,SeroneSpadaVilladoro2017}.
Related methods have been applied to two-dimensional scalar
field theory \cite{SeroneSpadaVilladoro2018}.  AUF instead works
with the original coefficient sequence, continued in complex order.

Consider the formal series
\begin{dmath}\label{eq:intro-formal-series}
 \widehat F(\zeta)={
 \sum_{n=0}^{\infty}\eta^n c_n\zeta^n,
 \qquad \eta\in\{-1,+1\}
 }
\end{dmath}.
AUF associates
the coefficients with an analytic function $c_F(t)$ satisfying
$c_F(n)=c_n$ for non-negative integers.  This is the
\emph{analytic coefficient clock}.  Integer interpolation alone
does not select it; its analytic domain and growth are part of
the construction. The clock is then paired with a fixed umbral kernel in a
Mellin--Barnes integral along a specified vertical contour.
Under the analyticity, growth and contour conditions stated below,
this pairing defines the AUF completion
$\widetilde F^{\AUF}(\zeta)$, with $\widehat F$ as its asymptotic
expansion on the corresponding sector as $\zeta\to0$.

For Gevrey-$s$ series, Watson-admissibility (W-admissibility)
controls both the Barnes integral and its strong Gevrey remainder.
The guaranteed sector has aperture $2\delta>\pi s$, and Watson
uniqueness identifies the completion with the canonical $k$-sum,
where $k=1/s$, in the specified directions.  With a common
admissible contour, gauge and logarithmic lift, admissible clocks
with the same samples also coincide on the connected common
domain containing that contour.  These restrictions distinguish
canonical completion from arbitrary interpolation or flat
additions on narrower sectors.

When $\delta>\pi$, adjacent determinations overlap on the
logarithmic covering.  Their clockwise difference gives
\begin{dmath}\label{eq:intro-new-hierarchy}
 \widehat F\longrightarrow c_F
 \overset{\rm W}{\longrightarrow}\widetilde F^{\AUF}
 \overset{\delta>\pi}{\longrightarrow}{
 \mathfrak M_F=2\pi\ii\widetilde J_F(\eta\zeta)
 }
\end{dmath}.
The lifted carrier $\widetilde J_F(\eta\zeta)$ is flat on the
overlap, with the logarithmic lift specified below.
It is itself the rational-gauge completion of a null clock,
whose integer samples vanish.
Its principal planar determination is $J_F\equiv J_{F,\pi}$.
The carrier collects the specified one-turn monodromy; it need
not separate individual Stokes contributions.
Boundary conditions, integration cycles or homogeneous solutions
can still provide independent data in a physical problem.

A second result extends the construction beyond this Gevrey
setting.  The lognormal series belongs to no finite Gevrey class,
so no finite-order Borel transform yields a convergent germ.
Nevertheless, Gaussian decay of its clock in imaginary order
gives an AUF completion and a nonzero ordinary flat carrier.
Selection requires explicit analytic conditions on the clock,
rather than Watson uniqueness.  Averaging a relaxation response
over lognormal rates supplies a physical realization of the same
coefficient growth.  This benchmark shows that AUF is more than
an alternative integral representation of finite-order
Borel--Laplace summation.

\vskip0.5cm
The article is organized as follows. We begin with Euler's equation in \cref{sec:euler}, so that the
construction and its meaning can be seen before the general
theory.  \Cref{sec:framework,sec:auf-monodromy,sec:auf-borel-laplace}
then establish canonical summation, the flat monodromy carrier
and the Borel--Laplace comparison.  The quartic and sourced-quartic
models, scaled Gamma function and solvable $O(4)$ toy model test
finite, ramified and infinite continuation, and an exact nonlinear
Stokes map.  \Cref{sec:ln-carriers} treats the non-Gevrey extension;
the discussion and conclusions assess the scope of the results.
Technical estimates are collected in the appendices.

\section{Euler's equation: completion and nonperturbative information
from one clock}
\label{sec:euler}

Euler's equation shows the construction in its simplest form.
We first obtain the completed perturbative solution from its
coefficient clock.  We then use the same clock to calculate an exact
exponentially small variation.  This example also distinguishes
information fixed by the perturbative completion from an arbitrary
homogeneous contribution.

Consider
\begin{dmath}\label{eq:euler-ode}
  g^2\frac{\dd E}{\dd g}+E(g)=g
\end{dmath},
whose formal solution is
\begin{dmath}\label{eq:euler-series}
  \widehat E(g)
  ={
  g\widehat H(g),
  \qquad
  \widehat H(g)
  =
  \sum_{n=0}^{\infty}(-1)^n n!g^n
  }
\end{dmath}.
Hats denote formal series.  We complete the normalized series
$\widehat H$ and recover $E$ by multiplication by $g$.
The normalized differential equation is
\begin{dmath}\label{eq:euler-H-ode}
  g^2H'(g)+(1+g)H(g)=1
\end{dmath}.

\subsection{Completing the perturbative solution}

The coefficients $n!$ have the analytic continuation
\begin{dmath}\label{eq:euler-clock}
  c_E(t)=\Gamma(1+t)
\end{dmath},
which we call the analytic coefficient clock.
Its recurrence $c_E(t+1)=(t+1)c_E(t)$ extends the factorial
recurrence from integer to complex order.
The subscript $E$ identifies the Euler example; the clock itself
belongs to the normalized series $\widehat H$.

The rational AUF gauge pairs this clock with a universal kernel.
Write $\ell=\ln|g|+\ii\vartheta$, where the lifted angle
$\vartheta\in\mathbb R$ records winding around the origin.
The completion is
\begin{dmath}\label{eq:euler-H-MB}
  \widetilde H^{\AUF}(\ell)
  ={
  -\frac{1}{2\pi\ii}
  \int_{\gamma-\ii\infty}^{\gamma+\ii\infty}
  \pi\ee^{t\ell}\csc(\pi t)\Gamma(1+t)\,\dd t,
  \qquad -1<\gamma<0,
  \quad |\IM\ell|<\frac{3\pi}{2}
  }
\end{dmath}.
The contour is oriented upward.  Displacing it to the right crosses
the poles at non-negative integers and reproduces the coefficients
$(-1)^n n!$.  The decay of the Gamma function also controls the
remainder, giving a strong Gevrey-one expansion on proper subsectors
of the stated domain.

The growth estimate is proved in \cref{app:euler-estimates}.
In the terminology developed below, its sharp Watson parameter is
$\delta_\ast=3\pi/2$.  The canonical-summation theorem then identifies
this completion with the directional Borel--Laplace sum for every
lifted direction $-\pi<d<\pi$.

We can evaluate the completion without first constructing a Borel
transform.  The clock has the moment representation
\begin{dmath}\label{eq:euler-weight}
  \rho_E(r)={
  \ee^{-r},
  \qquad
  c_E(t)=\int_0^\infty r^t\ee^{-r}\,\dd r
  }
\end{dmath}.
Substitution into the Barnes integral gives
\begin{dmath}\label{eq:euler-exact-integral}
  E^{\AUF}(g)
  ={
  g\int_0^\infty
  \frac{\ee^{-r}}{1+gr}\,\dd r,
  \qquad
  g\in\mathbb C^\ast\setminus L_\pi
  }
\end{dmath},
where $L_\pi$ is the negative real axis.
This planar determination uses the principal logarithm
$\Log g$, with $-\pi<\Arg g<\pi$.
Direct differentiation verifies \cref{eq:euler-ode}.
Equivalently,
\begin{dmath}\label{eq:euler-exact-E}
  E^{\AUF}(g)
  =
  \ee^{1/g}\operatorname E_1(1/g)
\end{dmath},
with the principal exponential integral.

For comparison, use the Borel convention
$\mathcal B(g^{n+1})=q^n/n!$.
The same coefficients give
\begin{dmath}\label{eq:euler-Borel}
  {
  \mathcal B\widehat E(q)=\frac{1}{1+q},
  \qquad
  \mathcal S_0^{\rm BL}\widehat{E}(g)
  =
  \int_0^\infty\frac{\ee^{-q/g}}{1+q}\,\dd q
  }
\end{dmath},
initially for positive $g$.
The substitution $q=gr$ recovers the preceding integral;
equality extends throughout the right half-plane.
Thus AUF and Borel--Laplace summation construct the same solution
here, using different integral representations.

\subsection{Recovering the carrier and monodromy}

To extract the one-turn variation, take the bare inverse Mellin
transform of the same clock on the logarithmic covering:
\begin{dmath}\label{eq:euler-carrier}
  \widetilde J_E(z)
  ={
  \frac{1}{2\pi\ii}
  \int_{\gamma-\ii\infty}^{\gamma+\ii\infty}
  z^t\Gamma(1+t)\,\dd t
  =
  \frac1z\exp\left(-\frac1z\right)
  }
\end{dmath}.
Here $z^t=\exp(t\log z)$ uses the covering logarithm.
The ordinary integral converges for $|\arg z|<\pi/2$.

Choosing $\Log z$ gives the principal planar carrier
$J_E\equiv J_{E,\pi}$.
Its explicit expression continues single-valuedly to
$\mathbb C^\ast$.
Its principal flat sector is $|\Arg z|<\pi/2$:
on each proper closed angular subsector there is a $c>0$ such that
\begin{dmath*}
  |J_E(z)|
  \leq
  |z|^{-1}\ee^{-c/|z|}
\end{dmath*}.
Continuation outside this sector does not preserve flatness.

A clockwise turn replaces $\ell$ by $\ell-2\pi\ii$.
Taking the difference of the two Barnes integrals cancels the
rational kernel and leaves the inverse Mellin integral:
\begin{dmath}\label{eq:euler-H-monodromy-carrier}
  \mathfrak M_H(g)
  ={
  \widetilde H^{\AUF}(\ell-2\pi\ii)
  -
  \widetilde H^{\AUF}(\ell)
  =
  2\pi\ii\widetilde J_E(-g)
  =
  -\frac{2\pi\ii}{g}\ee^{1/g}
  }
\end{dmath}.
The assigned carrier logarithm is $\log(-g)=\ell-\pi\ii$.
Both Barnes determinations coexist on
\begin{dmath}\label{eq:euler-flat-sector}
  \frac{\pi}{2}<{\arg g<\frac{3\pi}{2}}
\end{dmath},
where $\arg g$ denotes the lifted angle.
On this overlap, the carrier angle lies between $-\pi/2$ and
$\pi/2$, so the assigned logarithm also equals $\Log(-g)$.
Consequently $\widetilde J_E(-g)=J_E(-g)$ there, and the variation
can be represented by the principal planar carrier.
Multiplication by $g$ gives the variation of the original observable:
\begin{dmath}\label{eq:euler-clockwise-monodromy}
  \mathfrak M_E(g)
  ={
  g\mathfrak M_H(g)
  =
  -2\pi\ii\ee^{1/g}
  }
\end{dmath}.
The exponential contribution and its coefficient have therefore
been obtained directly from the perturbative clock.

The planar discontinuity is a boundary value of this analytic
variation.  We use increasing-angle ordering: on the negative axis
this means lower minus upper.  For $g=-a$, $a>0$,
\begin{dmath}\label{eq:euler-disc-carrier}
  \boxed{
  \Disc_\pi E_\pi^{\AUF}(-a)
  =
  -2\pi\ii\,aJ_E(a)
  =
  -2\pi\ii\ee^{-1/a}
  }
\end{dmath}.
In the real integral, the same difference comes from the pole at
$r=1/a$.  Deforming the integration contour allows continuation
through the negative axis on the logarithmic covering.
The cut specifies a planar determination; it is not a barrier
to continuation of the solution.

The equation permits continuation along every path in
$\mathbb C^\ast$.  Successive turns give distinct determinations,
as verified in \cref{app:euler-estimates}.
In this example the carrier consequently determines the entire
one-turn variation, with no further fluctuation series to resolve.

\subsection{What the perturbative completion fixes}

The exponential $\ee^{1/g}$ also solves the homogeneous Euler
equation and is flat on proper subsectors of
\cref{eq:euler-flat-sector}. With $g=\ee^\ell$, the lifted
completion of the original observable is
$\widetilde E^{\AUF}(\ell)=
\ee^\ell\widetilde H^{\AUF}(\ell)$.
Adding the homogeneous solution gives
\begin{dmath}\label{eq:euler-flat-family}
 \widetilde E_C(\ell)
 =\widetilde E^{\AUF}(\ell)+C\ee^{1/g}
\end{dmath}.
For every constant $C$, $\widetilde E_C$ has the same
perturbative expansion on those lifted subsectors.

This freedom does not contradict canonical summation.
The homogeneous addition grows as $g\to0^+$ and therefore fails
the asymptotic conditions selecting the canonical solution on its
wider sector.  A separate boundary-value problem may nevertheless
require such an addition and determine $C$ through global data.

Euler's example therefore separates two questions.
The clock fixes the completed perturbative solution and the
coefficient of its monodromy variation.
An independently chosen homogeneous contribution remains additional
information about the global solution.

\section{Analytic coefficient clocks and canonical AUF summation}
\label{sec:framework}

Euler's example suggests two requirements for the general construction.
The clock must interpolate the perturbative coefficients, and its growth
must allow contour displacement with a controlled remainder.  We now
state a sufficient condition that makes the resulting function the
canonical sum.  The monodromy carrier will be obtained from this same
clock in the next section.

\subsection{The clock, its gauges and logarithmic determinations}
\label{sec:analytic-clocks}

Consider
\begin{dmath}\label{eq:framework-formal-series}
 \widehat F(\zeta)={
 \sum_{n=0}^{\infty}\eta^n c_n\zeta^n,
 \qquad \eta\in\{-1,+1\}
 }
\end{dmath}.
The sign is a convention; the coefficients need not be positive or real.
An analytic coefficient clock $c_F(t)$ interpolates $c_F(n)=c_n$ for
non-negative integers.  Its analytic domain and growth conditions are
part of the construction.

Complex powers require a choice of logarithm. For the cut
$L_\phi=\{r\ee^{\ii\phi}:r>0\}$, we use
$\log_\phi\zeta=\ln|\zeta|+\ii\arg_\phi\zeta$, with
$\phi-2\pi<\arg_\phi\zeta<\phi$.
Thus $\Log=\log_\pi$ and $\Arg=\arg_\pi$ are principal,
while $\log_{\phi+2\pi}=\log_\phi+2\pi\ii$.
The unsubscripted symbols $\arg$ and $\log$ denote the angular
and logarithmic coordinates on the logarithmic covering, where
the angle is real and is not reduced modulo $2\pi$.
Restricting the angle to $\phi-2\pi<\vartheta<\phi$ selects a
portion of the covering that projects bijectively onto
$\mathbb C^\ast\setminus L_\phi$; $\arg_\phi$ and $\log_\phi$
are the corresponding planar representations.

The rational gauge uses the umbral functional
$f_\eta^{\rm R}(w)=(1-\eta w)^{-1}$, with $w=\zeta z$.
Its spectral kernel, defined by the Mellin transform \cite{RicciAUF}
\begin{dmath*}
	j_\eta^{\rm R}(t;\zeta)
	=
	\mathcal M_z\left[
	f_\eta^{\rm R}(\zeta z)
	\right](-t)
\end{dmath*},
contains a complex power of the transformed
argument $-\eta\zeta$.  We fix this power using the principal
logarithm $\Log(-\eta\zeta)$.

Set
\begin{dmath}\label{eq:eta-angles}
 {\theta_\eta=\frac\pi2(1-\eta),
 \qquad \chi_\eta=\frac\pi2(1+\eta)}
\end{dmath}.
The chosen logarithm places the reference cut in the
$\zeta$-plane at $L_{\theta_\eta}=\{\eta r:r>0\}$.
On this cut plane,
$\Log(-\eta\zeta)=\log_{\theta_\eta}\zeta+\ii\chi_\eta$.
Thus we use $\log_\pi\zeta$ for $\eta=-1$ and
$\log_0\zeta$ for $\eta=+1$.

With this determination, the Mellin kernel is
$j_\eta^{\rm R}(t;\zeta)
=-\pi\exp[t\Log(-\eta\zeta)]\csc(\pi t)$.
Pairing it with the clock defines the reference planar completion,
wherever the integral converges:
\begin{dmath}\label{eq:framework-rational-pairing}
 F^{\AUF}(\zeta)\equiv{
 F_{\theta_\eta}^{\AUF}(\zeta)
 =-\frac1{2\pi\ii}
 \int_{\mathcal C_\gamma}
 \pi\exp\left[t\Log(-\eta\zeta)\right]
 \csc(\pi t)c_F(t)\,\dd t}
\end{dmath},
where $\mathcal C_\gamma=\gamma+\ii\mathbb R$ is oriented upward,
with $-1<\gamma<0$ and the further restriction stated below.
The universal kernel supplies the perturbative poles; the clock
supplies their coefficients.

The reference cut fixes this planar representation.
A jump across it represents a directional Stokes jump only when
its boundary values are identified with the corresponding lateral
canonical sums.
On the logarithmic covering, write $\zeta=r\ee^{\ii\vartheta}$,
where $r>0$ and $\vartheta\in\mathbb R$ records winding around
the origin.  The assigned logarithms satisfy
\begin{dmath*}
 {
 \log\zeta=\ln r+\ii\vartheta,
 \qquad
 \log(-\eta\zeta)=\ln r+\ii(\vartheta+\chi_\eta)
 }
\end{dmath*}.
Replacing the principal logarithm in the pairing by the latter
expression defines the lifted function
$\widetilde F^{\AUF}(r,\vartheta)$. The angle $\vartheta$ records
the logarithmic sheet and is not identified modulo $2\pi$;
$\chi_\eta$ supplies the kernel's angular shift.
For Euler, $\chi_{-1}=0$.
We also write $\widetilde F^{\AUF}(\zeta)$ as shorthand for
$\widetilde F^{\AUF}(r,\vartheta)$, with the assigned logarithm
$\log\zeta=\ln r+\ii\vartheta$.
In this notation, the argument of a lifted function includes
its logarithmic sheet.

Choosing $\log_\phi$ gives the planar determination
\begin{dmath}\label{eq:lifted-planar-auf-relation}
 {F_\phi^{\AUF}(\zeta)
 =\widetilde F^{\AUF}(|\zeta|,\arg_\phi\zeta),
 \qquad |\arg_\phi\zeta+\chi_\eta|<\delta}
\end{dmath}.
Here $\zeta\in\mathbb C^\ast\setminus L_\phi$, and the angular
condition specifies the domain directly supplied by the Watson estimate.
Continuation beyond that domain requires separate justification.
The unsubscripted notation $F^{\AUF}$ denotes the reference
determination $F_{\theta_\eta}^{\AUF}$.
For an arbitrary cut, the adapted power is
$(-\eta\zeta)^t_{[\phi]}=
\exp[t(\log_\phi\zeta+\ii\chi_\eta)]$.

We order lateral differences by increasing angle:
\begin{dmath}\label{eq:theta-disc-definition}
 \Disc_\phi F(r\ee^{\ii\phi})
 =\lim_{\epsilon\to0^+}
 \left[
 F(r\ee^{\ii(\phi+\epsilon)})
 -F(r\ee^{\ii(\phi-\epsilon)})
 \right]
\end{dmath}.
This means upper minus lower on the positive axis and lower minus
upper on the negative axis.

The exponential gauge provides another factorization of the same
integrand.  The exponential0 umbral functional is $f_\eta^{\rm E}(w)=\ee^{\eta w}$;
its kernel and ground state are
\begin{dmath}\label{eq:framework-gamma-gauge}
 {
 j_\eta^{\rm E}(t;\zeta)=(-\eta\zeta)^t\Gamma(-t),
 \qquad
 \varphi_F(t)=\Gamma(1+t)c_F(t)
 }
\end{dmath}.
Euler's reflection formula gives
\begin{dmath}\label{eq:framework-gauge-integrand}
 j_\eta^{\rm E}(t;\zeta)\varphi_F(t)
 =j_\eta^{\rm R}(t;\zeta)c_F(t)
\end{dmath}.
Consequently the two gauges give the same completion on the same
contour and logarithmic determination.  The relation
$\varphi_F(n)=n!c_n$ also connects this factorization with the Mellin
interpolation underlying Ramanujan's Master Theorem
\cite{HardyRamanujan}.  Their integral relation to Borel--Laplace
summation is developed in \cref{sec:auf-borel-laplace}.

\subsection{Equations for the clock}
\label{sec:differential-difference}

A differential equation can help construct the clock, as Euler's
factorial recurrence illustrates.  Set
$\Theta_\zeta=\zeta\partial_\zeta$ and
$\mathcal T_t h(t)=h(t+1)$.  In the rational gauge,
\begin{dmath}\label{eq:framework-rational-transmutation}
 {
 \Theta_\zeta\mapsto t,
 \qquad
 \zeta\mapsto\eta\mathcal T_t^{-1}
 }
\end{dmath}.
For an operator ordered as
$\mathcal L(\zeta,\Theta_\zeta)=\sum_{j=0}^r\zeta^jP_j(\Theta_\zeta)$,
this gives the clock operator
\begin{dmath}\label{eq:framework-rational-operator}
 \mathcal L_{\rm R}(t,\mathcal T_t)
 =\sum_{j=0}^r\eta^j\mathcal T_t^{-j}P_j(t)
\end{dmath}.
Gamma conjugation gives the exponential-gauge rule
$\zeta\mapsto\eta t\mathcal T_t^{-1}$, with
$\Theta_\zeta\mapsto t$ unchanged.

These rules transport the Barnes contour along with the order variable.
Returning to a fixed contour requires convergence and any crossed
residues to be accounted for; such residues can supply an
inhomogeneous term.  An analytic recurrence is additional information
about the clock, not a consequence of assigning negative-index
coefficients.  The precise contour identity and low-order conditions
are given in \cref{app:auf-contour-transport}.

\subsection{Admissibility and canonical summation}
\label{sec:watson-clocks}

Interpolation alone does not ensure that the integral represents the
prescribed perturbative series.  The following sufficient condition
controls both the contour displacement and its remainder.
We call coefficient growth bounded by a constant times
$A^n\Gamma(1+sn)$ Gevrey-$s$; its summability order is $k=1/s$.

\begin{definition}[Watson-admissible clock]
\label{def:watson-clock}
Let $s>0$ and $\max\{-1,-1/s\}<\gamma<0$.
A clock is Watson-admissible if it is holomorphic in an open
neighborhood of $\RE t\geq\gamma$ and there are $C,A>0$, $m\geq0$
and $\delta>\pi s/2$ such that
\begin{dmath}\label{eq:framework-watson-bound}
 |c_F(\sigma+\ii\tau)|\leq{
 C A^\sigma\Gamma(1+s\sigma)
 (1+\sigma+|\tau|)^m
 \exp[(\pi-\delta)|\tau|],
 \qquad\sigma\geq\gamma,
 \quad\tau\in\mathbb R
 }
\end{dmath}.
Removable singularities are filled in before applying this condition.
\end{definition}

The Gamma factor controls growth in real order.  The imaginary-order
bound determines the angular domain: after multiplication by the kernel,
the net exponential decay is at least
$\exp[-(\delta-|\vartheta+\chi_\eta|)|\tau|]$.
The integral therefore converges on the lifted sector
\begin{dmath}\label{eq:framework-watson-angular-range}
 {|\vartheta+\chi_\eta|<\delta,\qquad\Omega_F=2\delta}
\end{dmath}.
Holomorphy ensures that rightward contour displacement crosses only
the perturbative poles.  Additional surviving poles would require a
different asymptotic projection, including their contributions.

\begin{theorem}[Canonical AUF summation]
\label{thm:mb-k-sum}
Let $c_F$ be Watson-admissible of Gevrey order $s=1/k$.
Then $\widetilde F^{\AUF}$ has $\widehat F$ as a strong Gevrey-$s$ expansion
on $|\vartheta+\chi_\eta|<\delta$.  More precisely, for every proper closed angular
subsector $S'$ and fixed $r_0>0$, there are constants independent
of $N$ and $\zeta$ such that
\begin{dmath}\label{eq:framework-gevrey-remainder}
 \left|\widetilde F^{\AUF}(\zeta)
 -\sum_{n=0}^{N-1}\eta^n c_n\zeta^n\right|
 \leq
 K_{S',r_0}B_{S',r_0}^N\Gamma(1+sN)|\zeta|^N
\end{dmath},
for all $N\geq0$ and $\zeta\in S'$ with $0<|\zeta|\leq r_0$.
The sum is empty for $N=0$.
For every lifted direction $d$ satisfying the condition below,
the series is $k$-summable and
\begin{dmath}\label{eq:framework-summable-directions}
 \boxed{{
 \widetilde F^{\AUF}=\mathcal S_{k,d}\widehat F,
 \qquad
 |d+\chi_\eta|<\delta-\frac{\pi s}{2}
 }}
\end{dmath},
on their connected common sectorial domain on the logarithmic
covering, with the same lift used for both functions.
\end{theorem}

The proof is in \cref{app:canonical-auf-proof}.
The remainder bound is stronger than coefficient growth alone:
it controls the completed function after every finite truncation.
Its sector is wide enough for Watson uniqueness, which identifies
the AUF function with the canonical $k$-sum \cite{Balser1994}.
Thus the clock constructs the function directly; summability theory
subsequently identifies it with the Borel--Laplace sum.

\begin{corollary}[Representative independence]
\label{cor:auf-representative-independence}
Two W-admissible clocks of the same Gevrey order with the same samples,
rational gauge, logarithmic lift and a common admissible vertical
contour give the same completion on
$|\vartheta+\chi_\eta|<\min\{\delta_1,\delta_2\}$.
Under these fixed-contour hypotheses they also coincide on the
connected common analytic domain containing the contour.
\end{corollary}

The first assertion follows from Watson uniqueness; the second follows
from Fourier uniqueness on the Barnes line, as detailed in
\cref{app:canonical-auf-proof}.
Arbitrary interpolations remain nonunique, but the stated admissibility
conditions select the canonical realization.
As Euler showed, flat additions on narrower sectors can still describe
independent global data.

Canonical summation completes the first step.  When $\delta>\pi$,
two adjacent logarithmic determinations overlap; the next section
calculates their difference from the same clock and establishes
the carrier's flatness on that overlap.

\section{AUF monodromy and the perturbatively encoded nonperturbative carrier}
\label{sec:auf-monodromy}

We now extract nonperturbative information from the completed function.
The calculation compares two determinations separated by one clockwise
turn around the origin.  Their difference removes the rational kernel
and leaves the bare inverse Mellin transform of the clock.
We then show that this exact difference has zero perturbative expansion
on the overlap of the two determinations.

\subsection{Adjacent determinations and the carrier}
\label{sec:monodromy-covering}

Retain the lifted argument $\vartheta=\arg\zeta$ introduced
in the preceding section.
The AUF integral converges for $|\vartheta+\chi_\eta|<\delta$.
Let $T$ increase $\vartheta$ by $2\pi$, so that
$(T^{-1}G)(r,\vartheta)=G(r,\vartheta-2\pi)$.
The clockwise monodromy variation is
\begin{dmath}\label{eq:monodromy-definition}
 \mathfrak M_F=(T^{-1}-1)\widetilde F^{\AUF}
\end{dmath}.
In the benchmarks, $M_E$, $M_0$ and $M_h$ denote counterclockwise
continuation operators: the realizations of $T$ on the corresponding
analytic germs. We reserve $\mathfrak M=T^{-1}-1$ for the clockwise
variation operator and $\mathfrak M_F$ for its value on the AUF
completion. Thus $M$ denotes continuation, whereas $\mathfrak M$
subtracts the initial determination after clockwise continuation.
Both determinations in \cref{eq:monodromy-definition} are supplied by the Barnes integral when
\begin{dmath*}
 {
 |\vartheta+\chi_\eta|<\delta,
 \qquad
 |\vartheta-2\pi+\chi_\eta|<\delta
 }
\end{dmath*}.
Their open overlap exists precisely when $\delta>\pi$.
Since $\theta_\eta+\chi_\eta=\pi$, it is
\begin{dmath}\label{eq:monodromy-sector}
 S_{\rm mon}
 ={
 \{(r,\vartheta):r>0,\ |\vartheta-\theta_\eta|<\delta-\pi\}
 }
\end{dmath}.
Its aperture is $2(\delta-\pi)$: the original Watson aperture with one
complete turn removed.  Canonical summation requires $\delta>\pi s/2$;
the ordinary carrier construction additionally requires $\delta>\pi$.

Define the bare inverse Mellin carrier on the logarithmic covering by
\begin{dmath}\label{eq:JF-definition}
 \widetilde J_F(z)=\frac1{2\pi\ii}
 \int_{\mathcal C_\gamma}
 \exp[t\log z]c_F(t)\,\dd t
\end{dmath}.
Here $z$ denotes a point on the covering.
We also write $\widetilde J_F(r,\alpha)$ for its value at
$\log z=\ln r+\ii\alpha$.

Removing the rational kernel removes its vertical damping
$\ee^{-\pi|\tau|}$.
For $\delta>\pi$, the remaining Watson bound guarantees absolute,
locally uniform convergence on
\begin{dmath}\label{eq:JF-sector}
 |\arg z|<\delta-\pi
\end{dmath}.

Choosing $\log_\phi$ gives the planar determination wherever
$|\arg_\phi z|<\delta-\pi$:
\begin{dmath*}
 {J_{F,\phi}(z)
 =\widetilde J_F(|z|,\arg_\phi z),
 \qquad J_F\equiv J_{F,\pi}}
\end{dmath*}.
Thus the unsubscripted carrier denotes its principal planar
determination, whose integral uses $\exp[t\Log z]$.
Its directly supplied domain is
\begin{dmath*}
 D_{F,\pi}
 ={
 \{z\in\mathbb C^\ast\setminus L_\pi:
 |\Arg z|<\delta-\pi\}
 }
\end{dmath*}.
Larger holomorphic or flat domains require separate justification.

For $z=\eta\zeta$ in the covering monodromy formula, the assigned
carrier logarithm is
\begin{dmath}\label{eq:carrier-logarithmic-lift}
 \log z=\ln r+\ii(\vartheta-\theta_\eta)
\end{dmath}.
The formula therefore uses $\widetilde J_F(\eta\zeta)$ with this lift.
On the portion of $S_{\rm mon}$ where
$-\pi<\vartheta-\theta_\eta<\pi$, this logarithm equals
$\Log(\eta\zeta)$, and the variation can equivalently be expressed
through the planar carrier $J_F(\eta\zeta)$.

\subsection{Exact monodromy and flatness}
\label{sec:monodromy-flatness-theorem}

The following theorem identifies the variation and establishes why
it is invisible to perturbation theory.  The same clock determines
both the completed function and this exact nonperturbative carrier.

\begin{theorem}[AUF monodromy-flatness theorem]
\label{thm:auf-monodromy-flatness}
Let $c_F$ be W-admissible of Gevrey order $s$, with $\delta>\pi$.
Then $\widetilde J_F$ is holomorphic on the covering sector defined
by \cref{eq:JF-sector}.
On $S_{\rm mon}$, with the carrier lift specified in
\cref{eq:carrier-logarithmic-lift},
\begin{dmath}\label{eq:exact-monodromy-JF}
 \boxed{\mathfrak M_F(\zeta)
 =2\pi\ii\widetilde J_F(\eta\zeta)}
\end{dmath}.
Moreover, on every proper closed angular subsector $S'$ of
$S_{\rm mon}$ there are $C_{S'},A_{S'},r_{S'}>0$ such that
\begin{dmath}\label{eq:JF-flat-estimate}
 |\widetilde J_F(\eta\zeta)|
 \leq C_{S'}A_{S'}^N\Gamma(1+sN)|\zeta|^N
\end{dmath},
for all $N\geq0$ and $\zeta\in S'$ with $0<|\zeta|<r_{S'}$.
The constants and radius are independent of $N$.
Consequently, the principal planar carrier $J_F$ is strongly
Gevrey-$s$ flat on every proper closed angular subsector of
$D_{F,\pi}$.
\end{theorem}

\begin{proof}
The two Barnes integrals converge absolutely on the overlap, so
their difference can be taken under the integral sign.
The identity
$(\ee^{-2\pi\ii t}-1)\pi\csc(\pi t)
=-2\pi\ii\ee^{-\pi\ii t}$ cancels the rational kernel and gives
\cref{eq:exact-monodromy-JF} with the lift
\cref{eq:carrier-logarithmic-lift}.

The two determinations have the same perturbative polynomial,
because integer powers are unchanged by a complete turn.
Subtract their strong Gevrey remainder bounds from
\cref{eq:framework-gevrey-remainder} and divide by $2\pi$.
Taking the larger exponential constant and the smaller radius gives
\cref{eq:JF-flat-estimate}.
Restricting the carrier angle to the principal interval gives
the stated planar flatness.
\end{proof}

The carrier also has a direct zero-series realization.  Define
\begin{dmath}\label{eq:null-clock}
 c_{J_F}(t)=-\frac{\sin(\pi t)}\pi c_F(t)
\end{dmath}.
All its non-negative integer samples vanish.
Its lifted rational-gauge completion, with sign $\eta=-1$, is
$\widetilde J_F$; its reference planar completion is $J_F$.
In both representations, the sine factor cancels the kernel.
Multiplication by the sine costs at most $\pi$ in the vertical
exponential estimate, leaving the inherited angular parameter
$\delta_J=\delta-\pi$.

A nonzero completion of these zero samples is consistent with
Watson uniqueness.
If $\delta_J>\pi s/2$, the null clock would itself satisfy the
strict Watson uniqueness condition, forcing $\widetilde J_F\equiv0$.
A nonzero carrier therefore requires
$\delta\leq\pi+\pi s/2$ for every Watson parameter of its original clock.

For fixed clock, Gevrey order and admissible half-plane, let
$\delta_\ast$ be the supremum of the admissible Watson parameters;
the bounding constants may depend on the parameter.
If $\delta_\ast>\pi$, the union of the guaranteed covering flat
sectors is
\begin{dmath}\label{eq:sharp-JF-sector}
 |\arg z|<\delta_\ast-\pi
\end{dmath}.
The supremum need not be attained: every proper subsector is
covered by an admissible parameter.
For the principal planar carrier, the corresponding domain is
\begin{dmath*}
 \{z\in\mathbb C^\ast\setminus L_\pi:
 |\Arg z|<\delta_\ast-\pi\}
\end{dmath*}.
These are W-guaranteed flat domains.
Larger flat domains require separate justification; analytic
continuation alone does not establish flatness.

\subsection{Planar jumps and the meaning of the carrier}
\label{sec:perturbatively-encoded-np}

The AUF carrier gives the \emph{aggregate, perturbatively encoded
nonperturbative variation} associated with one clockwise turn
of the selected completion.
Its interpretation in terms of directional Stokes variations
requires distinguishing planar cut jumps from comparisons
of lateral canonical sums.

A planar cut selects the logarithmic determinations used for
its two boundary values.  A directional Stokes variation compares
the lateral canonical sums of the perturbative series.
These pairs of functions need not coincide.

To express the covering variation in planar terms, set
$\mathcal U_\eta=\{\zeta\in\mathbb C^\ast:
\eta\zeta\in D_{F,\pi}\}$.
This sector contains the reference ray in its interior.
Let $F_+^{\AUF}$ and $F_-^{\AUF}$ denote the analytic
continuations into $\mathcal U_\eta$ of the two reference-bank
germs, with signs assigned by increasing-angle ordering.
For $\zeta\in\mathcal U_\eta$, put
$r=|\zeta|$ and $\vartheta=\theta_\eta+\Arg(\eta\zeta)$.
Their covering representations are
\begin{dmath*}
 {F_+^{\AUF}(\zeta)
 =\widetilde F^{\AUF}(r,\vartheta-2\pi),
 \qquad
 F_-^{\AUF}(\zeta)
 =\widetilde F^{\AUF}(r,\vartheta)}
\end{dmath*}.

Define the planar analytic monodromy variation by
\begin{dmath*}
 \bigl(\mathcal D_{\theta_\eta}^{\mathrm{mon}}
 F_{\theta_\eta}^{\AUF}\bigr)(\zeta)
 ={
 F_+^{\AUF}(\zeta)-F_-^{\AUF}(\zeta)
 =2\pi\ii J_F(\eta\zeta)
 }
\end{dmath*}.
The superscript $\mathrm{mon}$ specifies the continued
reference-bank pair, separated by one clockwise turn on the
covering.
In the Stokes benchmarks, $\mathcal D_\theta$ compares instead the
canonical sectorial sums selected on either side of direction
$\theta$, after placing them on a common domain.
Both notations abbreviate analytic differences between
specified pairs of functions.
For $\mathcal D_{\theta_\eta}^{\mathrm{mon}}$, the common domain
is $\mathcal U_\eta$, where the difference is holomorphic and
flat on proper closed angular subsectors as $\zeta\to0$.
The two comparisons agree when the continued reference-bank
germs are the corresponding lateral canonical sums.
The distinction between an analytic difference on a planar
overlap and its covering representation is illustrated explicitly
for Euler's series in
\cite[\S~10, Eqs.~(45)--(46)]{Sauzin2014}.

Restricting this analytic variation to the reference ray gives
\begin{dmath}\label{eq:natural-cut-disc-JF}
 \Disc_{\theta_\eta}F_{\theta_\eta}^{\AUF}
 (r\ee^{\ii\theta_\eta})
 = {
 \bigl(\mathcal D_{\theta_\eta}^{\mathrm{mon}}
 F_{\theta_\eta}^{\AUF}\bigr)(r\ee^{\ii\theta_\eta})
 =2\pi\ii J_F(r)
 }
\end{dmath}.
The ordering is lower minus upper for $\eta=-1$ and upper minus
lower for $\eta=+1$.
Thus $\Disc$ is the ray restriction of a sectorial analytic
variation.

When the continued bank germs coincide with the corresponding
lateral canonical sums, this comparison also gives the
directional Stokes variation.
The reference ray need not be a Stokes direction:
the lateral canonical sums may agree there even though the
two reference banks differ.
Conversely, a fixed planar determination may be holomorphic
across a Stokes ray where the lateral canonical sums differ;
its boundary values cannot then represent both canonical sums.
Other admissible cuts give other raywise restrictions of the
covering variation, as detailed in \cref{app:auf-planar-cuts}.

The word \emph{aggregate} refers to the specified one-turn
monodromy.
In the resurgent benchmarks, its exact expression is resolved
into the expected resummed nonperturbative contributions,
associated with possibly distinct Stokes directions.
A general decomposition into directional Stokes variations
is not assumed.

Under the ordinary inverse-Mellin hypotheses, Fourier uniqueness
on the Barnes line ensures that the principal carrier determines
the selected analytic clock.
In particular, $J_F\equiv0$ implies $c_F\equiv0$ and hence
$\widetilde F^{\AUF}\equiv0$.
AUF thus obtains an exact analytic nonperturbative variation
before any transseries decomposition.
Extracting actions, fluctuation series and directional Stokes
relations requires further analysis.
The qualification \emph{perturbatively encoded} remains essential:
independent homogeneous contributions or parameters supplied by
global conditions remain additional data.

A simple example, using two Euler components with non-collinear actions, helps clarify this point.
Take $\eta=+1$, $A=\ee^{\ii\alpha}$ and $\alpha=\pi/6$, with
$A^{-t}=\ee^{-\ii\alpha t}$:
\begin{dmath}\label{eq:two-action-euler-clock}
 {\widehat F_{1,A}(\zeta)
 =\sum_{n=0}^\infty n!(1+A^{-n})\zeta^n,
 \qquad
 c_{1,A}(t)=\Gamma(1+t)(1+A^{-t})}
\end{dmath}.
Since $|A^{-(\sigma+\ii\tau)}|=\ee^{\alpha\tau}$,
the Euler estimates in \cref{app:euler-estimates} give the sharp
symmetric Watson parameter $\delta_\ast=3\pi/2-\alpha=4\pi/3$.
The principal planar carrier is
\begin{dmath}\label{eq:two-action-euler-carrier}
 J_{1,A}(z)=\frac1z\ee^{-1/z}
 +\frac A z\ee^{-A/z}
\end{dmath}.
The ordinary inverse-Mellin integral converges absolutely, and
its value is flat, on
$-\pi/3<\Arg z<\pi/2$, uniformly on proper closed angular subsectors.
Both exponentials decay there.
The symmetric Watson bound guarantees the smaller sector
$|\Arg z|<\pi/3$; the explicit expression establishes the larger
flat sector and continues single-valuedly to $\mathbb C^\ast$.

Write $K_a(\zeta)=2\pi\ii(a/\zeta)\ee^{-a/\zeta}$ for
$a\in\{1,A\}$.
Applying Euler's lateral relation under the substitutions
$g=-\zeta$ and $g=-\zeta/A$ gives
\begin{dmath}\label{eq:two-action-euler-stokes}
 {(\mathcal S_{0+}-\mathcal S_{0-})\widehat F_{1,A}=K_1,
 \qquad
 (\mathcal S_{\alpha+}-\mathcal S_{\alpha-})
 \widehat F_{1,A}=K_A}
\end{dmath}.
Here $\mathcal S_{\theta\pm}$ denotes the lateral canonical
Gevrey-one sums (see next section), and each identity holds on its corresponding
sectorial overlap.

The reference completion $F_{1,A,0}^{\AUF}$ uses $\log_0$ and
satisfies
\begin{dmath}\label{eq:two-action-euler-reference-jumps}
 {\Disc_0F_{1,A,0}^{\AUF}(r)=K_1(r)+K_A(r),
 \qquad
 \Disc_\alpha F_{1,A,0}^{\AUF}(Ar)=0,
 \qquad r>0}
\end{dmath}.
At direction $0$, the directional Stokes variation is only $K_1$.
The additional term $K_A$ in the reference-cut jump is the
contribution associated with direction $\alpha$, evaluated on
the positive ray.

Near the second ray, the reference function agrees with
$\mathcal S_{\alpha+}\widehat F_{1,A}$ on both sides, while
$\mathcal S_{\alpha-}\widehat F_{1,A}
=F_{1,A,0}^{\AUF}-K_A$.
Its continuity through that ray therefore coexists with a
nonzero Stokes variation between the canonical sums.

If the action-$1$ component is omitted, direction $0$ is no
longer a Stokes direction, but the reference-cut jump remains
$K_A(r)$.
This isolates the distinction: a nonzero reference-cut jump
records monodromy even on a ray where the lateral canonical
sums coincide.

\section{Relation between AUF and Borel--Laplace summation}
\label{sec:auf-borel-laplace}

We now compare the AUF completion with Borel--Laplace summation.
For Gevrey-one series, the connection can be seen directly: a density
obtained from the carrier generates both a Cauchy representation of the
completed function and a Laplace representation of its Borel transform.
Inserting the latter into Borel--Laplace reconstruction recovers the
Cauchy integral.  We state the density assumptions explicitly; the
higher-Gevrey comparison then follows from canonical summation.

\subsection{The density and the two AUF gauges}
\label{sec:auf-cauchy-representation}

Assume an ordinary carrier and define its reciprocal density by
\begin{dmath}\label{eq:rho-J-relation}
 \rho_F(r)=\frac1r J_F\left(\frac1r\right)
\end{dmath}.
Suppose ordinary Mellin inversion gives
\begin{dmath}\label{eq:clock-moment-density}
 c_F(t)=\int_0^\infty r^t\rho_F(r)\,\dd r
\end{dmath},
absolutely on a common fundamental strip containing the Barnes line
$\RE t=\gamma$, with $-1<\gamma<0$.
In particular, $\int_0^\infty r^\gamma|\rho_F(r)|\,\dd r<\infty$.
Substitution into the rational pairing and absolute interchange give
\begin{dmath}\label{eq:rational-auf-cauchy-J}
 F^{\AUF}(\zeta)={
 \int_0^\infty\frac{\rho_F(r)}{1-\eta\zeta r}\,\dd r
 =\int_0^\infty\frac{J_F(z)}{z-\eta\zeta}\,\dd z
 }
\end{dmath},
on the reference cut plane $\mathbb C^\ast\setminus L_{\theta_\eta}$.
The second expression follows by $r=1/z$.
Its moving pole is $z=\eta\zeta$; the Plemelj boundary formula
therefore reproduces the carrier jump in
\cref{eq:natural-cut-disc-JF}, with the prescribed bank ordering.

The exponential gauge represents the same function with a different
density.  The ground state $\varphi_F(t)=\Gamma(1+t)c_F(t)$ has
the moment density
\begin{dmath}\label{eq:Phi-rho-mellin-convolution}
 \Phi_F(z)=\int_0^\infty
 \rho_F\left(\frac zu\right)\ee^{-u}\frac{\dd u}{u}
\end{dmath}.
Indeed, its Mellin moments are $\Gamma(1+t)c_F(t)$ on the common
strip restricted to $\RE t>-1$.
For $\RE(\eta\zeta)<0$, this gives
\begin{dmath}\label{eq:exponential-auf-laplace}
 F^{\AUF}(\zeta) = \int_0^\infty\ee^{\eta\zeta z}\Phi_F(z)\,\dd z
\end{dmath}.
The density hypotheses justify both gauge representations, but do not
yet identify a Borel germ.  The absolute-integration arguments are
collected in \cref{app:auf-bl-integrals}.

\subsection{The Borel germ and its reconstruction}
\label{sec:borel-germ-laplace-density}

For the Gevrey-one comparison, additionally assume
\begin{dmath}\label{eq:rho-exponential-moment}
 \int_0^\infty\ee^{\epsilon r}|\rho_F(r)|\,\dd r<\infty
\end{dmath},
for some $\epsilon>0$.
This controls all integer moments and permits integration of the
exponential series near the origin.
Use the shifted series $\widehat{\mathcal F}=\zeta\widehat F$ and
the convention $\mathcal B(\zeta^{n+1})=q^n/n!$.
Then
\begin{dmath}\label{eq:borel-laplace-rho}
 \mathcal B\widehat{\mathcal F}(q)={
 \sum_{n=0}^\infty\eta^n\frac{c_n}{n!}q^n
 =\int_0^\infty\ee^{\eta qr}\rho_F(r)\,\dd r
 }
\end{dmath}.
The equality follows by dominated integration for $|q|<\epsilon$;
the density integral continues the germ to $\RE(\eta q)<\epsilon$.

For its chosen determination along a lifted direction $d$, write
$\mathcal B_d\widehat{\mathcal F}$ and reconstruct by
\begin{dmath}\label{eq:directional-BL-sum}
 \mathcal S_d^{\rm BL}\widehat{\mathcal F}(\zeta)
 =\int_0^{\infty\ee^{\ii d}}
 \ee^{-q/\zeta}\mathcal B_d\widehat{\mathcal F}(q)\,\dd q
\end{dmath}.
An initial equality domain is obtained by choosing $d=-\chi_\eta$.
Since $\eta\ee^{\ii d}=-1$, the density formula represents the
Borel germ along the entire integration ray.
For $\RE(\eta\zeta)<0$, set
$h=\RE(\ee^{\ii d}/\zeta)>0$.
The absolute double integral is then
$\int_0^\infty|\rho_F(r)|/(h+r)\,\dd r<\infty$.
Fubini's theorem applies, and the inner Laplace integral becomes
$\zeta/(1-\eta\zeta r)$.
Comparison with \cref{eq:rational-auf-cauchy-J} gives
\begin{dmath}\label{eq:auf-bl-exact-identity}
 \boxed{
 \mathcal S_{-\chi_\eta}^{\rm BL}
 \widehat{\mathcal F}(\zeta)
 =\zeta F_{\theta_\eta}^{\AUF}(\zeta)
 }
\end{dmath},
on the half-plane $\RE(\eta\zeta)<0$.
Here $F_{\theta_\eta}^{\AUF}$ is the reference planar completion.

For other directions, the same integral reduction applies wherever
the density represents the chosen Borel continuation along the
whole contour and absolute interchange is justified.
Sufficient conditions are given in \cref{app:auf-bl-integrals}.
Thus the Borel exponential and the Cauchy kernel act on the same
density, establishing agreement on these specified domains.

Equality also extends along corresponding analytic continuations
of the common germ, with compatible logarithmic lifts.
The comparison must then use the corresponding AUF determination.
It does not identify every directional canonical sum with the
fixed reference planar function: on a shared asymptotic sector,
they can differ by a flat contribution.
The two-action Euler example in
\cref{eq:two-action-euler-stokes,eq:two-action-euler-reference-jumps}
illustrates this distinction.

\subsection{Higher Gevrey order}
\label{sec:higher-gevrey-bl}

For Gevrey-$s$ series, set $k=1/s$ and use the unshifted transform
\begin{dmath}\label{eq:order-k-borel}
 \mathcal B_k\widehat F(\xi)
 =\sum_{n=0}^\infty\eta^n
 \frac{c_n}{\Gamma(1+n/k)}\xi^n
\end{dmath}.
For a continued determination $b=\mathcal B_{k,d}\widehat F$,
the inverse reconstruction is
\begin{dmath}\label{eq:order-k-laplace}
 \mathcal L_{k,d}\,b(\zeta)
 =k\zeta^{-k}\int_0^{\infty\ee^{\ii d}}
 \exp[-(\xi/\zeta)^k]b(\xi)\xi^{k-1}\,\dd\xi
\end{dmath}.
All powers use compatible logarithmic lifts, with
$\log(\xi/\zeta)=\log\xi-\log\zeta$.
The reconstruction is taken on the central component
$|\vartheta-d|<\pi/(2k)$.
If $|b(\xi)|\leq C\exp(K|\xi|^k)$ along the ray, convergence holds
there when $\RE(\ee^{\ii kd}\zeta^{-k})>K$, hence for sufficiently
small $|\zeta|$ on proper subsectors.
The normalization is
$\mathcal L_{k,d}[\xi^n]=\Gamma(1+n/k)\zeta^n$.

Under the summability hypotheses,
$\mathcal S_{k,d}\widehat F
=\mathcal L_{k,d}\mathcal B_{k,d}\widehat F$.
For the directions admitted by
\cref{eq:framework-summable-directions},
\cref{thm:mb-k-sum} identifies this reconstruction with
$\widetilde F^{\AUF}$ on their connected common sectorial domain
on the logarithmic covering, using the same assigned logarithm
on both sides.
At $k=1$, $\mathcal B_1\widehat F$ equals
$\mathcal B\widehat{\mathcal F}$, while the unshifted reconstruction
has the prefactor $\zeta^{-1}$, consistently with
\cref{eq:auf-bl-exact-identity}.
The explicit density reduction above concerns Gevrey one;
its extension to general $k$ requires a separately justified kernel
factorization.

\vskip0.5cm
We now apply the construction to examples with progressively richer
nonperturbative structure.  In each benchmark, the clock first determines
the completed function and its aggregate carrier; exact formulas or
differential equations then resolve that carrier into actions,
fluctuation series and continuation data.  Keeping these two steps
distinct shows what follows from AUF completion itself and what requires
further analysis.  The final lognormal example tests the construction
beyond finite Gevrey order.

\section{The symmetric quartic model: flat carrier and finite monodromy}
\label{sec:quartic}

This model shows how an exact carrier yields an instanton action and
its fluctuation series.  It also shows why the number of determinations
need not equal the dimension of the solution space: this scalar solution
has four determinations, all contained in a two-dimensional solution space.

For $\RE x<0$, use the normalization \cite{Aniceto2019}
\begin{dmath}\label{eq:quartic-partition}
 Z_0(x)={
 \frac{\sqrt{\hbar}}{2\pi}\int_{\mathbb R}
 \exp\left(-\frac{z^2}{2}+\frac{xz^4}{24}\right)\,\dd z,
 \qquad \hbar>0
 }
\end{dmath}.
Gaussian moments give
\begin{dmath}\label{eq:quartic-perturbative-series}
 \widehat Z_0(x)={
 Z_0^{(0)}\sum_{n=0}^\infty a_nx^n,
 \qquad a_n=\frac{\Gamma(2n+\frac12)}
 {6^n\sqrt{\pi}\Gamma(1+n)},
 \qquad Z_0^{(0)}=\sqrt{\frac{\hbar}{2\pi}}
 }
\end{dmath}.
Thus $\eta=+1$.  Duplication writes the clock as
\begin{dmath}\label{eq:quartic-clock}
 c_0(t)={
 Z_0^{(0)}6^{-t}\frac{\Gamma(2t+\frac12)}
 {\sqrt{\pi}\Gamma(1+t)}
 =\frac{\sqrt{\hbar}}{2\pi^{3/2}}
 \left(\frac23\right)^t
 \frac{\Gamma(t+\frac14)\Gamma(t+\frac34)}{\Gamma(1+t)}
 }
\end{dmath}.
The recurrence
$c_0(t+1)=\frac{2}{3}(t+\frac14)(t+\frac34)c_0(t)/(t+1)$
leads, with the contour-transport justification of
\cref{app:canonical-auf-proof}, to
\begin{dmath}\label{eq:quartic-ode}
 16x^2Z''(x)+(32x-24)Z'(x)+3Z(x)=0
\end{dmath}.

\subsection{Canonical completion and its Watson sector}

Choose $-\frac14<\gamma<0$.
The beta integral bounds
$|\Gamma(t+\frac34)/\Gamma(t+1)|$ by the same ratio at
$t=\RE t=\sigma$.
Together with the rotated Gamma estimate from the Euler benchmark,
this gives, for every $0<\beta<\pi/2$,
\begin{dmath}\label{eq:quartic-watson-bound}
 |c_0(\sigma+\ii\tau)|\leq{
 C_\beta A_\beta^\sigma\Gamma(1+\sigma)
 \ee^{-\beta|\tau|},\qquad \sigma\geq\gamma
 }
\end{dmath}.
Hence the clock is W-admissible of Gevrey order one with
$\delta=\pi+\beta$.
At fixed $\sigma$, Stirling's formula gives
$|c_0(\sigma+\ii\tau)|\sim
C_\sigma|\tau|^{\sigma-1/2}\ee^{-\pi|\tau|/2}$,
with $C_\sigma>0$, so $\delta_\ast=3\pi/2$ is sharp.

Write $\vartheta=\arg x$ for the lifted argument and assign
$L=\ln|x|+\ii(\vartheta+\pi)$ to $\log(-x)$.
The rational-gauge completion is
\begin{dmath}\label{eq:quartic-rational-mb}
 \widetilde Z_0^{\AUF}(L)
 =-\frac1{2\pi\ii}\int_{\mathcal C_\gamma}
 \pi\ee^{tL}\csc(\pi t)c_0(t)\,\dd t
\end{dmath}.
It converges on $|\vartheta+\pi|<3\pi/2$.
The smaller Watson parameters establish the strong Gevrey expansion on
every proper subsector and canonical summability in the lifted
directions $-2\pi<d<0$.
The reference planar determination uses
$L=\Log(-x)=\log_0x+\pi\ii$ on
$\mathbb C^\ast\setminus L_0$.

To identify it with the original integral, use the exponential-gauge
density
\begin{dmath}\label{eq:quartic-exponential-density}
 \Phi_0(v)=\frac{\sqrt{\hbar}\,24^{1/4}}{4\pi}
 v^{-3/4}\ee^{-\sqrt{6v}}
\end{dmath}.
Its moments are $\Gamma(1+t)c_0(t)$, and the substitution $v=z^4/24$
shows that $\int_0^\infty\ee^{xv}\Phi_0(v)\,\dd v$ equals
\cref{eq:quartic-partition} for $\RE x<0$.
Continuation of that germ yields the exact formula
\begin{dmath}\label{eq:quartic-exact-K}
 Z_0^{\AUF}(x)=\frac{\sqrt{3\hbar}}{2\pi\sqrt{-x}}
 \ee^{-3/(4x)}K_{1/4}\left(-\frac{3}{4x}\right)
\end{dmath},
with the square root and Bessel function taken on the determination
specified by $L$.

\subsection{The carrier and its instanton fluctuations}

For this clock, \cref{eq:JF-definition} defines the lifted carrier
$\widetilde J_0$.
Its principal planar determination $J_0\equiv J_{0,\pi}$ is
\begin{dmath}\label{eq:quartic-carrier}
 J_0(x)=\frac{\sqrt{\hbar}}{2\pi^2}\sqrt{\frac3{2x}}
 \ee^{-3/(4x)}K_{1/4}\left(\frac3{4x}\right)
\end{dmath}.
The square root uses $\Log x$, and the Bessel function uses its
principal branch.
This expression is holomorphic on $\mathbb C^\ast\setminus L_\pi$.
The ordinary inverse-Mellin integral converges initially on
$|\Arg x|<\pi/2$, where the carrier is strongly Gevrey-one flat
on proper closed angular subsectors.

To check the normalization, set $y=3/(4x)$ and use
\begin{dmath}\label{eq:quartic-K-Mellin}
 \int_0^\infty y^{t-1/2}\ee^{-y}K_{1/4}(y)\,\dd y
 =\frac{\sqrt{\pi}\Gamma(t+\frac14)\Gamma(t+\frac34)}
 {2^{t+1/2}\Gamma(1+t)}
\end{dmath},
for $\RE t>-\frac14$.
The resulting Mellin moments reproduce $c_0(t)$.
The null clock is $c_{J_0}(t)=-\sin(\pi t)c_0(t)/\pi$;
its inherited angular parameters approach the Watson uniqueness
threshold from below, as in Euler's example.

Define the exact second solution
\begin{dmath}\label{eq:quartic-Z1-definition}
 Z_1(x)=-\frac{\ii}{\sqrt2}\ee^{-3/(2x)}Z_0^{\AUF}(-x)
\end{dmath},
using the reference planar determination of $Z_0^{\AUF}(-x)$.
Here $\log_0(-x)=\Log x-\pi\ii$, so both Bessel expressions use
compatible principal branches.
Their comparison gives $J_0=\ii Z_1/\pi$ throughout
$\mathbb C^\ast\setminus L_\pi$.

On the adjacent-sheet overlap $-\pi/2<\vartheta<\pi/2$,
the assigned carrier logarithm $L-\pi\ii$ equals $\Log x$.
Hence the clockwise variation is
$\mathfrak M_0=2\pi\ii\widetilde J_0(x)
=2\pi\ii J_0(x)=-2Z_1(x)$ there.
On the reference cut this becomes
\begin{dmath}\label{eq:quartic-disc-exact}
 \Disc_0Z_0^{\AUF}(r)={
 Z_0^{\rm up}(r)-Z_0^{\rm low}(r)
 =2\pi\ii J_0(r)=-2Z_1(r),\qquad r>0
 }
\end{dmath},
in agreement with the resurgent normalization of \cite{Aniceto2019}.

The large-argument expansion of the carrier's Bessel factor gives
\begin{dmath}\label{eq:quartic-carrier-asymptotic}
 J_0(x)\sim\frac{Z_0^{(0)}}{\sqrt2\pi}\ee^{-3/(2x)}
 \sum_{n=0}^\infty(-1)^na_nx^n
\end{dmath},
on proper subsectors of $|\Arg x|<\pi/2$.
Thus the clock-derived carrier supplies the action $A=3/2$ and
its divergent fluctuation series.
The exact function $Z_1$ is already available before that series
is separately summed.
The reciprocal density $\rho_0(r)=r^{-1}J_0(1/r)$ also satisfies
the Borel comparison hypotheses: it is $\mathcal O(r^{-3/4})$
near zero and $\mathcal O(r^{-1}\ee^{-3r/2})$ at infinity,
so it has absolute exponential moments for $0<\epsilon<3/2$.

\subsection{Four determinations in a two-dimensional solution space}

The differential equation permits continuation along every path in
$\mathbb C^\ast$.  At a reference point $r>0$, choose the upper-bank
germ $Z_0^{\rm up}$ and the initial germ of $Z_1$.
Their sheets matter: the upper bank has $L=\ln r-\pi\ii$ and
the lower bank $L=\ln r+\pi\ii$.
A counterclockwise turn $M_0$ therefore gives
$M_0Z_0^{\rm up}=Z_0^{\rm low}=Z_0^{\rm up}+2Z_1$.

To determine subsequent turns, set $y=-3/(4x)$ with
$\log y=\ln(3/4)-L$.
The Bessel solution space has basis
$\ee^y y^{1/2}I_{-1/4}(y)$ and
$\ee^y y^{1/2}I_{1/4}(y)$.
A counterclockwise turn of $x$ decreases $\log y$ by $2\pi\ii$;
the basis functions acquire multipliers $-\ii$ and $\ii$.
Hence $M_0^2=-I$ and $M_0^4=I$.
Together with the first variation, this fixes
\begin{dmath}\label{eq:quartic-monodromy-matrix}
 M_0\begin{pmatrix}Z_0^{\rm up}\\Z_1\end{pmatrix}
 =\begin{pmatrix}1&2\\-1&-1\end{pmatrix}
 \begin{pmatrix}Z_0^{\rm up}\\Z_1\end{pmatrix}
\end{dmath}.
The two germs are independent: $Z_0^{\rm up}$ has the nonzero
perturbative constant $Z_0^{(0)}$, whereas $Z_1$ is flat.
The scalar germ therefore has four distinct determinations,
\begin{dmath}\label{eq:quartic-Z0-orbit}
 \left\{Z_0^{\rm up},\ Z_0^{\rm up}+2Z_1,
 -Z_0^{\rm up},\ -Z_0^{\rm up}-2Z_1\right\}
\end{dmath},
and maximal continuation surface
\begin{dmath}\label{eq:quartic-maximal-surface}
 X_0\simeq{
 \widetilde{\mathbb C^\ast}/4\mathbb Z,\qquad x=w^4
 }
\end{dmath}.
This is an unramified four-sheeted covering of the punctured plane
\cite{Forster1981}.

Continuation also changes the carrier.
Using the Bessel continuation of $\widetilde J_0$, define the
planar determinations represented on $\mathbb C^\ast\setminus L_\pi$ by
\begin{dmath*}
 {J_0^{[m]}(x)
 =\widetilde J_0(|x|,\Arg x+2\pi m),
 \qquad m\in\mathbb Z}
\end{dmath*}.
Here $J_0^{[0]}=J_0$, and their germs at the reference point
$r>0$ satisfy $J_0^{[m]}=M_0^mJ_0$.
The successive increments are
$M_0^{m+1}Z_0^{\rm up}-M_0^mZ_0^{\rm up}
=-2\pi\ii J_0^{[m]}$.

The monodromy matrix gives
$J_0^{[1]}=-\ii Z_0^{\rm up}/\pi-J_0$ and
$J_0^{[m+2]}=-J_0^{[m]}$.
Consequently, the even determinations are flat on
$|\Arg x|<\pi/2$, whereas the odd determinations have a nonzero
perturbative component there.
The four successive increments sum to zero.
Thus AUF supplies the first monodromy difference directly, while
the differential equation and Bessel analysis establish its finite
continuation structure.

\section{The sourced quartic model: source-resolved carriers}
\label{sec:sourced-quartic}

A source separates the two nonperturbative contributions that coincide
in the symmetric model.  We first obtain their exact carriers from
half-line clocks, then extract the different exponential scales by
steepest descent.  The total function retains four determinations.

Consider the normalized integral
\begin{dmath}\label{eq:sourced-partition}
 \mathcal Z(\lambda,h)=\frac{\ee^{-h^2/2}}{\sqrt{2\pi}}
 \int_{\mathbb R}
 \exp\left(-\frac{q^2}{2}-hq-\frac{\lambda q^4}{4}\right)
 \,\dd q
\end{dmath},
for $\RE\lambda>0$.
The source $h$ is fixed, with $\mathcal Z(0,h)=1$.
The constructions are holomorphic in $h$, with bounds locally uniform
on compact source sets.

\subsection{Half-line clocks and canonical completion}

The expansion is
$\widehat{\mathcal Z}(\lambda,h)=\sum_{n\geq0}(-1)^nc_n(h)\lambda^n$,
so $\eta=-1$.
Splitting the Gaussian moments gives
\begin{dmath}\label{eq:sourced-clock-components}
 c_{h,\sigma}(t)={
 \frac{4^{-t}\ee^{-h^2/2}}{\sqrt{2\pi}\Gamma(1+t)}
 \int_0^\infty q^{4t}\ee^{-q^2/2-\sigma hq}\,\dd q,
 \qquad \sigma=\pm1
 }
\end{dmath},
for $\RE t>-1/4$.
The total clock is $c_h=c_{h,+}+c_{h,-}$, with $c_h(n)=c_n(h)$.

Fix $-1/4<\gamma<0$ and write $t=a+\ii\tau$.
Rotating the $q$-ray through $\operatorname{sgn}(\tau)\phi$,
where $0<\phi<\pi/4$, contributes $\ee^{-4\phi|\tau|}$.
Absorbing the source into a Gaussian majorant and bounding
$\Gamma(1+t)^{-1}$ gives
\begin{dmath}\label{eq:sourced-watson-bound}
 |c_{h,\sigma}(a+\ii\tau)|\leq{
 C A^a\Gamma(1+a)(1+a+|\tau|)^m
 \ee^{(\pi/2-4\phi)|\tau|},\qquad a\geq\gamma
 }
\end{dmath}.
Each component is W-admissible of Gevrey order one with
$\delta=\pi/2+4\phi$.
The union of these estimates guarantees the lifted completion sector
$|\vartheta|<3\pi/2$ and carrier sector $|\arg u|<\pi/2$,
with strong Gevrey estimates on proper subsectors.

The rational pairing with $c_h$ defines $\mathcal Z^{\AUF}$.
In the exponential gauge, the moment integral for
$\Gamma(1+t)c_{h,\sigma}(t)$ reconstructs the corresponding
half-line integral, initially for $\RE\lambda>0$.
Thus $\mathcal Z^{\AUF}=\mathcal Z_+^{\AUF}+\mathcal Z_-^{\AUF}
=\mathcal Z$ there, and along compatible continuations.

\subsection{Exact carriers and their source-dependent scales}

Define the lifted inverse Mellin carriers by
\begin{dmath}\label{eq:sourced-carriers}
 \widetilde J_{h,\sigma}(u)={
 \frac1{2\pi\ii}\int_{\mathcal C_\gamma}
 u^t c_{h,\sigma}(t)\,\dd t,
 \qquad
 \widetilde J_h=\widetilde J_{h,+}+\widetilde J_{h,-}
 }
\end{dmath}.
Here $u^t=\exp(t\log u)$ uses the covering logarithm.
The ordinary integrals converge on $|\arg u|<\pi/2$.
Their principal planar determinations are
$J_{h,\sigma}\equiv J_{h,\sigma,\pi}$ and
$J_h=J_{h,+}+J_{h,-}$.
Each is strongly Gevrey-one flat on $|\Arg u|<\pi/2$,
uniformly on proper closed angular subsectors and compact source sets.

Expanding the source exponential and setting $z=1/(4u)$ gives
\begin{dmath}\label{eq:sourced-carrier-explicit}
 J_{h,\sigma}(u)=\frac{\ee^{-h^2/2-z}}{2\pi\sqrt2}
 \sum_{m=0}^\infty\frac{(-2\sigma h)^m}{m!}
 z^{m/4+3/4}
 U\left(\frac34-\frac m4,\frac32,z\right)
\end{dmath},
where $U$ is the confluent hypergeometric function on its principal
branch and
$\Log z=-\ln4-\Log u$ fixes all powers.
Specializing the Laplace-transform identity in
\cite[Eq.~13.10.7]{NIST:DLMF} at unit Laplace parameter verifies
the inversion term by term.
The rotated-contour majorant justifies the exchange locally
uniformly in $h$ and on proper subsectors of the initial carrier sector.
The global continuation established in
\cref{eq:sourced-global-lift,eq:sourced-component-monodromy} below
extends these principal carriers holomorphically to
$\mathbb C^\ast\setminus L_\pi$.

On the reference negative-axis cut, the exact lower-minus-upper
jump is
\begin{dmath}\label{eq:sourced-disc-decomposition}
 \Disc_\pi\mathcal Z_\pi^{\AUF}(-\rho,h)={
 2\pi\ii\left[J_{h,+}(\rho)+J_{h,-}(\rho)\right],
 \qquad\rho>0
 }
\end{dmath}.
The two contributions are therefore separated before any saddle
expansion is performed.

The saddle normalization follows from the contour form of each
half-line discontinuity:
\begin{dmath}\label{eq:sourced-carrier-contour}
 J_{h,\sigma}(\rho)=
 \frac{\ee^{-h^2/2}}{2\pi\ii\sqrt{2\pi}}
 \int_{\mathcal C_{\rm jump}}
 \exp\left(-\frac{q^2}{2}-\sigma hq+\frac{\rho q^4}{4}\right)
 \,\dd q
\end{dmath}.
Here $\mathcal C_{\rm jump}$ runs from infinity on the ray
$\arg q=-\pi/4$ to the origin and then to infinity on
$\arg q=\pi/4$.  This orientation represents lower minus upper:
continuation to the lower bank rotates the original half-line
upward, whereas continuation to the upper bank rotates it downward.

For fixed real $h$ and sufficiently small $\rho>0$, let $q_\sigma$
be the large positive root of
$\rho q_\sigma^3-q_\sigma-\sigma h=0$.  Then
\begin{dmath}\label{eq:sourced-q-asymptotic}
 q_\sigma=\rho^{-1/2}+\frac{\sigma h}{2}
 -\frac{3h^2}{8}\rho^{1/2}+\mathcal O(\rho)
\end{dmath}.
The integrand is entire, so the contour can be deformed to
$q(y)=\sqrt{q_\sigma^2+y^2}+\ii y$, with $y$ increasing from
$-\infty$ to $+\infty$.  It passes upward through $q_\sigma$
and approaches the same decay rays.  Along this contour, the real
part of the exponent is bounded above by its saddle value minus
$(cy^2+\rho y^4)$, with $c>0$ for sufficiently small $\rho$.
The saddle action is
\begin{dmath}\label{eq:sourced-action-expansion}
 \mathcal S_\sigma={
 \frac{q_\sigma^2}{2}+\sigma hq_\sigma
 -\frac{\rho q_\sigma^4}{4}
 =\frac1{4\rho}+\frac{\sigma h}{\sqrt\rho}
 +\frac{h^2}{4}+\mathcal O(\sqrt\rho)
 }
\end{dmath}.
The quadratic action coefficient at the saddle is
$1-3\rho q_\sigma^2=-2+\mathcal O(\sqrt\rho)$.
Since $\dd q=\ii\,\dd y$ there, the leading Gaussian factor is
$\ii\sqrt{\pi}$.  Combining it with the prefactor in
\cref{eq:sourced-carrier-contour} yields
\begin{dmath}\label{eq:sourced-carrier-asymptotic}
 J_{h,\sigma}(\rho)=\frac1{2\pi\sqrt2}
 \exp\left[-\frac1{4\rho}-\frac{\sigma h}{\sqrt\rho}
 -\frac{3h^2}{4}\right]
 \left[1+\mathcal O(\sqrt\rho)\right]
\end{dmath},
as $\rho\to0^+$.
In the carrier variable $u=-\lambda$, the common leading action
is $1/4$; along $u=\rho>0$, the source distinguishes the two scales
$\exp(\mp h/\sqrt\rho)$.
The fluctuation expansions use powers of $\sqrt\rho$, and the two
formal families are exchanged when its determination changes sign.
This describes their asymptotic resolution, rather than a change in
the number of sheets of the total function.
At $h=0$, the components coincide and
\begin{dmath}\label{eq:sourced-symmetric-limit}
 {\mathcal Z(\lambda,0)=\frac{Z_0^{\AUF}(-6\lambda)}{Z_0^{(0)}},
 \qquad J_h(u)\big|_{h=0}=\frac{J_0(6u)}{Z_0^{(0)}}}
\end{dmath}.

\subsection{Continuation of the exact functions}

Assign $L=\ln|\lambda|+\ii\vartheta$ to $\log\lambda$.
Scaling $q=\ee^{-L/4}y$ gives the global lift
\begin{dmath}\label{eq:sourced-global-lift}
 \widetilde{\mathcal Z}(L,h)=\frac{\ee^{-h^2/2}}{\sqrt{2\pi}}
 \ee^{-L/4}\int_{\mathbb R}
 \exp\left[-\frac{y^4}{4}-\frac12\ee^{-L/2}y^2
 -h\ee^{-L/4}y\right]\,\dd y
\end{dmath}.
It agrees with the original integral for real $L$.
Quartic damping supplies locally uniform domination on compact
sets of $(L,h)$, so this expression is entire in $L$ and continues
the AUF germ.
The half-line lifts $\widetilde{\mathcal Z}_\sigma(L,h)$ are
obtained by integrating over $y>0$ and replacing the source
$h$ by $\sigma h$. Their sum is $\widetilde{\mathcal Z}(L,h)$.

Let $M_h$ act on these lifted functions by
$(M_hG)(L,h)=G(L+2\pi\ii,h)$.
After two turns the prefactor changes sign and the half-line source
is reversed, giving
$M_h^2\widetilde{\mathcal Z}_\sigma
=-\widetilde{\mathcal Z}_{-\sigma}$.
Their sum therefore satisfies
\begin{dmath}\label{eq:sourced-monodromy-square}
 {M_h^2\widetilde{\mathcal Z}=-\widetilde{\mathcal Z},
 \qquad
 M_h^4\widetilde{\mathcal Z}=\widetilde{\mathcal Z}}
\end{dmath}.
Since the total germ is nonzero, its stabilizer is exactly
$4\mathbb Z$ and its maximal continuation surface is
$X_h\simeq\widetilde{\mathbb C^\ast}/4\mathbb Z$
\cite{Forster1981}.

On the initial overlap $-3\pi/2<\vartheta<-\pi/2$,
the counterclockwise increment is
\begin{dmath}\label{eq:sourced-component-monodromy}
 (M_h-I)\widetilde{\mathcal Z}_\sigma(L,h)
 =-2\pi\ii\widetilde J_{h,\sigma}
 \left(\ee^{L+\pi\ii}\right)
\end{dmath},
with assigned carrier logarithm $L+\pi\ii$.
The left-hand side is entire in $L$.
Solving this identity for the carrier therefore supplies its
unique continuation as an entire function of the logarithmic
coordinate $L+\pi\ii$, and the identity holds for all $L$
after this continuation.
Its principal planar restrictions are consequently holomorphic
on $\mathbb C^\ast\setminus L_\pi$.
Their W-guaranteed flat sector remains $|\Arg u|<\pi/2$.

On the initial overlap, the carrier angle lies between
$-\pi/2$ and $\pi/2$, so
$L+\pi\ii=\Log(-\lambda)$.
The right-hand side can therefore also be written as
$-2\pi\ii J_{h,\sigma}(-\lambda)$.
At $L=\ln\rho-\pi\ii$, the turn goes from the lower to the upper
bank, giving the negative of the jump convention in
\cref{eq:sourced-disc-decomposition}.
The source thus resolves the first monodromy variation into two
flat carriers while the total completed function retains four sheets.

\section{The scaled Gamma function:
a countable tower from one carrier}
\label{sec:gamma-model}

Here one inverse Mellin carrier contains a countable tower of
exponential contributions.  We construct it in the parent variable
$x=g^2$, where the series is Gevrey two, then translate its monodromy
into the physical variable $g$, where the Stirling series is Gevrey one.
The factor $g$ and the logarithmic lift must both be retained in this
translation.

Consider
\begin{dmath}\label{eq:gamma-scaled}
 Z(g)={\Gstar(1/g),\qquad
 \Gstar(z)=\frac{\Gamma(z)}
 {\sqrt{2\pi}\,z^{z-1/2}\ee^{-z}}}
\end{dmath},
and set $W(g)=\log Z(g)$, with powers and logarithms initially fixed
by positive real arguments.
Its Stirling expansion is \cite{NIST:DLMF,MitschiSauzin2016}
\begin{dmath}\label{eq:gamma-Stirling}
 \widehat W(g)=\sum_{n\geq0}
 \frac{B_{2n+2}}{(2n+2)(2n+1)}g^{2n+1}
\end{dmath},
where $B_N$ are Bernoulli numbers.
Factoring out $g$ gives
\begin{dmath}\label{eq:gamma-parent-series}
 \widehat W(g)={g\widehat H(g^2),\qquad
 \widehat H(x)=\sum_{n=0}^\infty(-1)^nc_\Gamma(n)x^n}
\end{dmath},
with $\eta=-1$ and clock
\begin{dmath}\label{eq:gamma-clock}
 c_\Gamma(t)=\frac{2\Gamma(1+2t)}{(2\pi)^{2t+2}}
 \zeta(2t+2)
\end{dmath}.
Here $\zeta$ is the Riemann zeta function.
Its Dirichlet series will resolve the carrier into individual terms.

\subsection{Canonical completion and Binet's formula}

Choose $-1/2<\gamma<0$.
The absolutely convergent Dirichlet series bounds the zeta factor
uniformly for $\RE t\geq\gamma$.
A rotated Gamma integral then gives, for every $0<\beta<\pi/2$,
\begin{dmath}\label{eq:gamma-watson-bound}
 |c_\Gamma(\sigma+\ii\tau)|\leq{
 C_\beta A_\beta^\sigma\Gamma(1+2\sigma)
 \ee^{-2\beta|\tau|},\qquad\sigma\geq\gamma}
\end{dmath}.
Thus the clock is W-admissible of Gevrey order two with
$\delta=\pi+2\beta$.
On a fixed vertical line, Stirling's formula gives Gamma decay
$\ee^{-\pi|\tau|}$ up to powers of $|\tau|$.
Since $\zeta$ and $1/\zeta$ are bounded in this Dirichlet half-plane,
the sharp value is $\delta_\ast=2\pi$.

Write $L=\log x$ for the assigned logarithm of the parent variable.
The lifted completion is
\begin{dmath}\label{eq:gamma-rational-mb}
 \widetilde H^{\AUF}(L)
 =-\frac1{2\pi\ii}\int_{\mathcal C_\gamma}
 \pi\ee^{tL}\csc(\pi t)c_\Gamma(t)\,\dd t
\end{dmath}.
It converges on $|\IM L|<2\pi$; the smaller Watson parameters establish
the strong Gevrey-two expansion on every proper subsector.
Its reference planar determination is
$H^{\AUF}(x)=\widetilde H^{\AUF}(\Log x)$ on
$\mathbb C^\ast\setminus L_\pi$, so $H^{\AUF}=H_\pi^{\AUF}$.

To identify this function, use the moment density
\begin{dmath}\label{eq:gamma-weight}
 {w_\Gamma(s)=-\frac{\ln(1-\ee^{-2\pi\sqrt s})}{2\pi\sqrt s},
 \qquad c_\Gamma(t)=\int_0^\infty s^tw_\Gamma(s)\,\dd s}
\end{dmath},
for $s>0$ and $\RE t>-1/2$.
Its behavior $s^{-1/2}|\ln s|$ near zero and
$s^{-1/2}\ee^{-2\pi\sqrt s}$ at infinity justifies the Cauchy
reduction on the reference parent cut plane.
This density has no absolute exponential moment in $s$;
the Gevrey-one density comparison is therefore not applied to the
parent series.

On the physical determination centered at positive $g$, set
$W_0(g)=g\widetilde H^{\AUF}(2\Log g)$.
For $\RE g>0$, one has $\Log(g^2)=2\Log g$, and hence
$W_0(g)=gH^{\AUF}(g^2)$.
Setting $s=v^2$ in the Cauchy representation therefore gives
\begin{dmath}\label{eq:gamma-binet-form}
 W_0(g)=-\frac g\pi\int_0^\infty
 \frac{\ln(1-\ee^{-2\pi v})}{1+g^2v^2}\,\dd v
\end{dmath},
initially for $\RE g>0$.
Integration by parts, with vanishing endpoint terms, yields
\begin{dmath}\label{eq:gamma-binet-second}
 W_0(g)=2\int_0^\infty
 \frac{\arctan(gv)}{\ee^{2\pi v}-1}\,\dd v
\end{dmath}.
This is Binet's second formula for $\log\Gamma(z)$ after subtracting
the elementary Stirling terms and setting $z=1/g$
\cite[Eq.~5.9.10]{NIST:DLMF}.
Hence $W_0(g)=\log\Gstar(1/g)$ on the determination fixed by
positive $g$.
The lifted AUF integral continues it to $|\vartheta|<\pi$, where
$\vartheta=\IM\log g$.
Substitution of $x=g^2$ and multiplication by $g$ convert the parent
remainder estimates into strong Gevrey-one bounds for $W_0$.

\subsection{The regular carrier and its exponential tower}

The lifted carrier $\widetilde J_\Gamma$ of
\cref{eq:JF-definition} converges on $|\arg u|<\pi$.
Its principal determination $J_\Gamma\equiv J_{\Gamma,\pi}$
is holomorphic on $\mathbb C^\ast\setminus L_\pi$ and strongly
Gevrey-two flat on proper closed angular subsectors.
Termwise inversion of the Dirichlet series gives
\begin{dmath}\label{eq:gamma-carrier-tower}
 J_\Gamma(u)=\frac1{2\pi\sqrt u}
 \sum_{k=1}^\infty\frac1k\ee^{-2\pi k/\sqrt u}
\end{dmath},
with $\sqrt u=\exp(\frac12\Log u)$.
Absolute Dirichlet convergence and vertical Gamma decay justify
the exchange locally uniformly.
Since $\RE(1/\sqrt u)>0$, the series sums to
\begin{dmath}\label{eq:gamma-carrier}
 J_\Gamma(u)=-\frac1{2\pi\sqrt u}
 \log\left(1-\ee^{-2\pi/\sqrt u}\right)
\end{dmath},
where the logarithm is fixed by the convergent series.
The reciprocal relation is $w_\Gamma(s)=s^{-1}J_\Gamma(1/s)$.

The null clock $-\sin(\pi t)c_\Gamma(t)/\pi$ completes to
$\widetilde J_\Gamma$ on the cover and $J_\Gamma$ on the principal
plane; its inherited angular parameters approach the Gevrey-two
Watson threshold $\pi$ from below.
The parent jump is
$\Disc_\pi H_\pi^{\AUF}(-u)=2\pi\ii J_\Gamma(u)$ for $u>0$,
with lower minus upper ordering.

\subsection{Physical Stokes variations and action resolution}

The Binet poles $v=\pm\ii/g$ meet the positive integration ray at
$\vartheta=\pm\pi/2$; both physical Stokes directions project onto
the negative parent axis.
Using the parent lift in \cref{eq:gamma-rational-mb}, define
the physical sectorial sums by
\begin{dmath}\label{eq:gamma-sectorial-sums}
 W_d(g)={g\widetilde H^{\AUF}(2\log g-2\ii d),
 \qquad d\in\{0,\pi,-\pi\}}
\end{dmath}.
They have the same formal expansion, each on $|\vartheta-d|<\pi$.
The Stokes variation compares these selected sums, even though each
continues analytically across the comparison ray.

On the upper overlap $0<\vartheta<\pi$, the increasing-angle
variation is
\begin{dmath}\label{eq:gamma-carrier-stokes}
 \mathcal D_{\pi/2}W(g)={
 W_\pi(g)-W_0(g)=2\pi\ii gJ_\Gamma(-g^2)}
\end{dmath},
with $\log(-g^2)=2\log g-\pi\ii$ and $\sqrt{-g^2}=-\ii g$.
Substitution gives both the exact variation and its convergent
action resolution:
\begin{dmath}\label{eq:gamma-analytic-variation}
 \mathcal D_{\pi/2}W(g)={
 \log\left(1-\ee^{-2\pi\ii/g}\right)
 =-\sum_{k=1}^\infty\frac1k\ee^{-2\pi\ii k/g}}
\end{dmath}.
The logarithm is fixed by this series.
On $g=\ii a$, $a>0$, the variation is
$\ln(1-\ee^{-2\pi/a})$.
It is strongly Gevrey-one flat on proper upper-half-plane subsectors,
with actions $A_k=2\pi\ii k$, $k\geq1$.

On the lower overlap $-\pi<\vartheta<0$, the same ordering gives
\begin{dmath}\label{eq:gamma-lower-variation}
 \mathcal D_{-\pi/2}W(g)={
 W_0(g)-W_{-\pi}(g)
 =-\log\left(1-\ee^{2\pi\ii/g}\right)}
\end{dmath},
with carrier lift $\log(-g^2)=2\log g+\pi\ii$ and
$\sqrt{-g^2}=\ii g$.
Here too the logarithm is selected by its convergent exponential series.

The action lattice matches the translated Gamma saddles:
for $S(\phi)=\ee^\phi-\phi-1$, the saddles
$\phi_k=2\pi\ii k$ have actions $-2\pi\ii k$.
Translation preserves the local fluctuation problem; the upper
suppressed family uses negative saddle indices
\cite{SauzinGamma2021}.
The weights $1/k$ belong to the logarithmic observable.
With $Z_d=\ee^{W_d}$, exponentiating gives
\begin{dmath}\label{eq:gamma-Z-ratio}
 \frac{Z_\pi(g)}{Z_0(g)}=1-\ee^{-2\pi\ii/g}
\end{dmath}.
Thus the countable tower and the multiplicative Stokes relation both
follow from one exact clock-derived carrier.

\section{A solvable \texorpdfstring{$O(4)$}{O(4)} toy model:
carrier-generated analytic closure}
\label{sec:o4-toy}

The $O(4)$ toy model of \cite[Appendix~G]{BajnokBalogVona2022} shows how
an infinite nonlinear transseries can arise from two exact functions.
The carrier first supplies a second solution of the perturbative
solution's differential equation.  Their ratio then obeys a
fractional-linear Stokes transformation whose expansion generates
the nonlinear hierarchy.

With $\eta=+1$, the perturbative series and its clock are
\begin{dmath}\label{eq:o4-toy-clock}
 {\widehat B(x)=\sum_{n=0}^\infty c_B(n)x^n,
 \qquad c_B(t)=\frac{\Gamma(t+\frac12)^2}{\pi\Gamma(1+t)}}
\end{dmath}.
The clock is holomorphic for $\RE t>-1/2$ and satisfies
$c_B(t+1)=(t+\frac12)^2c_B(t)/(t+1)$.
With the contour-transport justification of
\cref{app:canonical-auf-proof}, this recurrence induces
\begin{dmath}\label{eq:o4-toy-B-ode}
 {\left[\Theta_x-x\left(\Theta_x+\frac12\right)^2\right]B=0,
 \qquad\Theta_x=x\frac{\dd}{\dd x}}
\end{dmath}.

\subsection{Canonical completion and the sharp Watson sector}

Fix $-1/2<\gamma<0$.
The beta representation of $\Gamma(t+\frac12)/\Gamma(1+t)$,
combined with the rotated Gamma estimate, gives for $0<\beta<\pi/2$
\begin{dmath}\label{eq:o4-toy-watson-bound}
 |c_B(\sigma+\ii\tau)|\leq{
 C_{\gamma,\beta}(\sec\beta)^\sigma\Gamma(1+\sigma)
 \ee^{-\beta|\tau|},\qquad\sigma\geq\gamma}
\end{dmath}.
Thus the clock is W-admissible of Gevrey order one with
$\delta=\pi+\beta$.
At fixed $\gamma$, Stirling's formula gives
$|c_B(\gamma+\ii\tau)|\sim
\sqrt{2/\pi}\,|\tau|^{\gamma-1/2}\ee^{-\pi|\tau|/2}$,
so $\delta_\ast=3\pi/2$ is sharp.

Write $\ell=\ln|x|+\ii\vartheta$ on the logarithmic covering.
The rational-gauge completion is
\begin{dmath}\label{eq:o4-toy-B-MB}
 \widetilde B^{\AUF}(\ell)
 =-\frac1{2\pi\ii}\int_{\mathcal C_\gamma}
 \pi\ee^{t(\ell+\pi\ii)}\csc(\pi t)c_B(t)\,\dd t
\end{dmath}.
Its guaranteed sector is $|\vartheta+\pi|<3\pi/2$, with strong
Gevrey-one asymptotics on proper subsectors and canonical summation
directions $-2\pi<d<0$.
The reference planar determination $B$ uses $\ell=\log_0x$ and is
analytic on $\mathbb C^\ast\setminus L_0$.
An Euler integral for one Gamma factor in the pairing gives
\begin{dmath}\label{eq:o4-toy-integral}
 B(x)=\frac1{\sqrt{\pi}}\int_0^\infty
 \frac{\ee^{-s}}{\sqrt{s}\sqrt{1-xs}}\,\dd s
\end{dmath},
with the square root continued from $x<0$.
Evaluation yields
\begin{dmath}\label{eq:o4-toy-B-exact}
 B(x)={
 \left(-\frac1x\right)^{1/2}U\left(\frac12,1,-\frac1x\right)
 =\frac1{\sqrt{\pi}}\left(-\frac1x\right)^{1/2}
 \ee^{-1/(2x)}K_0\left(-\frac1{2x}\right)}
\end{dmath},
where $\log(-1/x)=-\log_0x-\pi\ii$.
The Tricomi--Bessel identity is
\cite[Eq.~13.6.10]{NIST:DLMF}; the lifted expressions use the same
assigned logarithm.

For comparison, in the shifted-series convention of
\cref{sec:auf-borel-laplace}, the Borel germ is
\begin{dmath}\label{eq:o4-toy-borel}
 \mathcal B[x\widehat B](q)={
 \sum_{n=0}^\infty\frac{c_B(n)}{n!}q^n
 ={}_2F_1\left(\frac12,\frac12;1;q\right)}
\end{dmath}.
Its directional Laplace reconstruction gives $xB(x)$ on the
corresponding common domain, identifying the same canonical function.

\subsection{The carrier as a second solution}

The principal determination $J_B\equiv J_{B,\pi}$ of the lifted
carrier $\widetilde J_B$ is
\begin{dmath}\label{eq:o4-toy-carrier}
 J_B(u)=\frac{\ee^{-1/u}}\pi B(-u)
\end{dmath},
with $\log_0(-u)=\Log u-\pi\ii$.
It is holomorphic on $\mathbb C^\ast\setminus L_\pi$.
The ordinary inverse-Mellin integral converges on $|\Arg u|<\pi/2$,
where $J_B$ is strongly Gevrey-one flat on proper closed angular
subsectors.
Its Mellin moments reproduce $c_B(t)$ for $\RE t>-1/2$, using
the Euler integral for $U(\frac12,1,1/u)$.
Its null-clock realization follows from \cref{eq:null-clock}.

Introduce
\begin{dmath}\label{eq:o4-toy-D-C}
 {D(x)=B(-x),\qquad C(x)=\ee^{-1/x}D(x)=\pi J_B(x)}
\end{dmath}.
The identity holds on $\mathbb C^\ast\setminus L_\pi$.
Substitution into \cref{eq:o4-toy-B-ode} shows that $C$ is a
second solution, whereas $D$ alone is not.
Choosing either lateral germ of $B$ near the positive ray,
$\{B,C\}$ forms a local basis: $B$ has leading term $1$ while
$C$ is exponentially flat.
The carrier has action $A=1$ and fluctuation expansion
$D(x)\sim\sum_{n\geq0}(-1)^nc_B(n)x^n$.

On the covering overlap $|\vartheta|<\pi/2$, the assigned carrier
logarithm $\ell$ equals $\Log x$, and
$(T^{-1}-1)\widetilde B^{\AUF}
=2\pi\ii\widetilde J_B(x)=2\pi\ii J_B(x)=2\ii C(x)$.
On the reference planar cuts, the raywise relations are
\begin{dmath}\label{eq:o4-toy-raywise-jumps}
 {\Disc_0B(r)=2\ii C(r),\qquad
 \Disc_\pi C(-r)=2\ii B(-r),\qquad r>0}
\end{dmath}.
The first is upper minus lower.
For the second, lower minus upper for $D(x)=B(-x)$ gives
$\Disc_\pi D(-r)=2\ii\ee^{-1/r}B(-r)$;
multiplication by $\ee^{-1/x}$ cancels that exponential.
Their composition requires tracking the logarithmic determinations.
The stated planar flat sector is $|\Arg x|<\pi/2$.

\subsection{Infinite continuation and the nonlinear observable}

The differential equation permits continuation along every path in
$\mathbb C^\ast$.
Fix $\ell=\log_0x$ for the base planar determination and set
$y=-1/(2x)$, with $\log y=-\ell-\ln2-\pi\ii$.
After $n$ counterclockwise turns of $x$, $y$ makes $n$ clockwise
turns.  The Bessel connection formula
\cite[Eq.~10.34.5]{NIST:DLMF} gives
\begin{dmath}\label{eq:o4-toy-monodromy-n}
 B^{[n]}(x)=(-1)^n\sqrt{\frac2\pi}\,y^{1/2}\ee^y
 \left[K_0(y)+2\pi\ii nI_0(y)\right]
\end{dmath},
for $n\in\mathbb Z$.
Independence of $K_0$ and $I_0$ excludes any nonzero winding that
returns the initial germ.
Hence the maximal continuation surface is
$X_B\simeq\widetilde{\mathbb C^\ast}$ \cite{Forster1981}.
Its infinitely many scalar determinations remain in the same
two-dimensional solution space.

The nonlinear observable of
\cite[Appendix~G, Eq.~G.30]{BajnokBalogVona2022} is
\begin{dmath}\label{eq:o4-toy-epsilon}
 \epsilon(x)={
 \frac{B(x)}{D(x)}
 =\ee^{-1/x}\frac{B(x)}{C(x)}
 }
\end{dmath},
where the denominator is nonzero.
For $r>0$, put $q=\ee^{-1/r}$ and let the subscripts
$\theta\pm$ denote cut-adapted lateral values at arguments $\theta\pm0$.
On the positive ray $D$ has no jump; on the negative ray $B$ has
no jump and $D_{\pi+}=D_{\pi-}+2\ii qB$.
The two exact transformations therefore are
\begin{dmath}\label{eq:o4-toy-epsilon-Stokes}
 {\epsilon_{0+}=\epsilon_{0-}+2\ii q,\qquad
 \epsilon_{\pi+}=\frac{\epsilon_{\pi-}}
 {1+2\ii q\epsilon_{\pi-}}}
\end{dmath}.
The plus sign in the denominator follows from the same
increasing-angle convention on both rays.

The second relation is a M{\"o}bius transformation.
For $|2q\epsilon_{\pi-}|<1$, expansion of its denominator gives
\begin{dmath}\label{eq:o4-toy-negative-expansion}
 \epsilon_{\pi+}=\epsilon_{\pi-}
 -2\ii q\epsilon_{\pi-}^2
 +(2\ii q)^2\epsilon_{\pi-}^3-\cdots
\end{dmath}.
Near the negative ray, these powers of $q$ organize the exponentially
small corrections, with fluctuation series supplied by the powers
of $\epsilon_{\pi-}$.
The infinite nonlinear hierarchy thus follows from an exact
transformation of the carrier-generated pair $\{B,C\}$.
The solution-space dimension, the number of continuation sheets and
the number of transseries terms are three distinct features.

\section{Beyond finite Gevrey order: lognormal completion and its carrier}
\label{sec:ln-carriers}

We now test the AUF construction on a non-Gevrey series: one lying
outside every finite Gevrey class.  The preceding benchmarks admit canonical
Borel--Laplace sums; the lognormal series below has no convergent
Borel germ of any finite order.  Nevertheless, its clock gives a
convergent AUF completion and an ordinary, nonzero flat carrier.
This tests the scope of the analytic pairing beyond the Watson
hypothesis.  We establish ordinary asymptoticity separately and
specify the additional conditions that select the clock.

Throughout, $\eta=-1$, contours point upward, and powers use assigned
logarithms.  We retain $\mathfrak M=T^{-1}-1$, with $T$ increasing
the logarithm by $2\pi\ii$.

\subsection{Completion without a finite Gevrey bound}

Gaussian decay in imaginary order can ensure convergence even when
real-order growth exceeds every factorial majorant.

\begin{proposition}[Gaussian decay in complex order]
\label{prop:ln-gaussian-class}
Let $c$ be holomorphic near $\RE t\geq\gamma$, with $-1<\gamma<0$.
Suppose every compact interval $I\subset[\gamma,\infty)$ admits
constants $C_I,B_I,a_I>0$ and $m_I\geq0$ such that
\begin{dmath}\label{eq:ln-gaussian-strip}
 |c(\sigma+\ii\tau)|\leq{
 C_I(1+|\tau|)^{m_I}\exp(-a_I\tau^2+B_I|\tau|),
 \qquad\sigma\in I}
\end{dmath}.
Then the lifted rational pairing associated with
\cref{eq:framework-rational-pairing} and the inverse Mellin
integral \cref{eq:JF-definition} define
$\widetilde F^{\AUF}$ and $\widetilde J_F$,
holomorphic on the whole logarithmic covering.
Their ordinary asymptotic expansions (regular projections) are
$\sum_{n\geq0}(-1)^nc(n)\zeta^n$ and the zero series, respectively,
on every bounded interval of lifted arguments.
The carrier integral is independent of its abscissa $d\geq\gamma$, and
\begin{dmath}\label{eq:ln-general-monodromy}
 \mathfrak M\widetilde F^{\AUF}(\zeta)={
 2\pi\ii\widetilde J_F(-\zeta),\qquad
 \log(-\zeta)=\log\zeta-\pi\ii}
\end{dmath}.
A nonzero clock has a nonzero carrier.
The principal planar carrier $J_F\equiv J_{F,\pi}$ is holomorphic
on $\mathbb C^\ast\setminus L_\pi$ and flat on every proper closed
angular subsector of that cut plane.
\end{proposition}

The proof, by finite contour displacement, is in
\cref{app:non-gevrey-estimates}.
No uniform Gevrey bound as $\RE t\to\infty$ is required.
The null clock $-\sin(\pi t)c(t)/\pi$ satisfies the same strip
condition and completes the zero series to $\widetilde J_F$
by kernel cancellation.

\subsection{The lognormal family and clock selection}

For $a>0$ and $b\in\mathbb R$, take
$\widehat F_{a,b}(\zeta)=\sum_{n\geq0}(-1)^nM_n^{(a,b)}\zeta^n$, with
\begin{dmath}\label{eq:ln-formal-data}
 {M_n^{(a,b)}=\frac{\ee^{a n^2/2+b n}}{\Gamma(1+n)},
 \qquad c_{a,b}(t)=\frac{\ee^{a t^2/2+b t}}{\Gamma(1+t)}}
\end{dmath}.
Since $\ln M_n^{(a,b)}=a n^2/2+O(n\ln n)$, division by
$\Gamma(1+s n)$ leaves zero convergence radius for every finite
$s>0$.  Thus ordinary and finite-order Borel--Laplace summation
cannot construct this series.
The clock does satisfy \cref{prop:ln-gaussian-class}, as do finite
positive mixtures of these clocks.
For $q=\ee^a$, its coefficient growth is of $q$-Gevrey type;
comparison with specialized $q$-summation or acceleration remains
possible \cite{LastraSanzSendra2022}.

Selection requires more than integer samples.
Require an entire clock with the complex-order recurrence
\begin{dmath}\label{eq:ln-clock-recurrence}
 (t+1)c(t+1)={\ee^{a t+a/2+b}c(t),\qquad c(0)=1}
\end{dmath},
whose multiplier
$p_c(t)=\ee^{-a t^2/2-bt}\Gamma(1+t)c(t)$ extends to an entire
function satisfying
$|p_c(x+\ii y)|\leq C\ee^{(2\pi-\epsilon)|y|}$ for
$0\leq x\leq1$, with $0<\epsilon<2\pi$.
The recurrence makes $p_c$ periodic with period $1$.
Vertical translation of its Fourier integrals eliminates every
nonconstant mode under this strict type bound.
Hence $p_c=1$ and $c=c_{a,b}$.
These are explicit selection conditions, not consequences of the samples.

The strict threshold matters even for positive moment measures.
Consider
\begin{equation}\label{eq:ln-boundary-measures}
 \begin{aligned}
 \dd\mu_{a,b}(x)&=
 \frac1{\sqrt{2\pi a}}
 \exp\left[-\frac{(\ln x-b)^2}{2a}\right]\frac{\dd x}{x},\\
 \dd\mu_\varepsilon(x)&=
 \left[1+\varepsilon\sin\left(\frac{2\pi(\ln x-b)}a\right)\right]
 \dd\mu_{a,b}(x),\qquad \varepsilon\in[-1,1].
 \end{aligned}
\end{equation}.
Their complex moments are
\begin{dmath}\label{eq:ln-boundary-moments}
 \int_0^\infty x^t\,\dd\mu_\varepsilon(x)
 =\ee^{a t^2/2+bt}
 \left[1+\varepsilon\ee^{-2\pi^2/a}\sin(2\pi t)\right]
\end{dmath}.
Thus distinct probability measures have identical integer moments.
Their clocks obey the same recurrence, but nonconstant multipliers
have vertical type $2\pi$.
The resulting distinct completions share the same ordinary expansion;
positivity and the recurrence do not remove this ambiguity.

For the selected clock, the lifted completion is
\begin{dmath}\label{eq:ln-canonical-completion}
 \widetilde F_{a,b}^{\AUF}(\zeta)=\frac1{2\pi\ii}
 \int_{\mathcal C_\gamma}\ee^{t\log\zeta}\Gamma(-t)
 \ee^{a t^2/2+bt}\,\dd t
\end{dmath},
with $-1<\gamma<0$, where $\log\zeta$ is the logarithmic
coordinate on the covering.
We write $F_{a,b}^{\AUF}=F_{a,b,\pi}^{\AUF}$ for its reference
planar determination, obtained by using $\Log\zeta$ in this integral.
Inserting Gaussian moments into the Mellin representation of the
exponential gives, for $\RE\zeta>0$,
\begin{dmath}\label{eq:ln-moment-realization}
 F_{a,b}^{\AUF}(\zeta)=\int_0^\infty
 \ee^{-\zeta x}\,\dd\mu_{a,b}(x)
\end{dmath}.
Finite negative moments and vertical Gamma decay justify the interchange
in this half-plane. This Mellin--Barnes representation was previously
obtained by Miles \cite{Miles2022}.
The last integral is the probability Laplace transform of the moment
measure; it does not require a Borel transform of the formal series.
The finite exponential remainder gives
\begin{dmath}\label{eq:ln-boundary-remainder}
 \left|F_{a,b}^{\AUF}(\zeta)
 -\sum_{n=0}^{N-1}(-1)^nM_n^{(a,b)}\zeta^n\right|
 \leq{M_N^{(a,b)}|\zeta|^N,\qquad\RE\zeta\geq0,\quad N\geq1}
\end{dmath}.
The same bound holds for the other positive measures above.
Canonical selection here refers to the specified clock class.

\subsection{The ordinary carrier and its flat scale}

The selected lifted carrier is
\begin{dmath}\label{eq:ln-carrier-definition}
 \widetilde J_{a,b}(z)=\frac1{2\pi\ii}
 \int_{d+\ii\mathbb R}z^t
 \frac{\ee^{a t^2/2+bt}}{\Gamma(1+t)}\,\dd t
\end{dmath}.
Here $z^t=\exp(t\log z)$.
The entire clock allows any real $d$, and Gaussian decay gives
an entire function of $\log z$, flat on every bounded lifted
angular interval.
Its principal determination $J_{a,b}\equiv J_{a,b,\pi}$ is
holomorphic on $\mathbb C^\ast\setminus L_\pi$ and flat on proper
closed angular subsectors.
The reference-cut jump is
$\Disc_\pi F_{a,b,\pi}^{\AUF}(-r)=2\pi\ii J_{a,b}(r)$,
with lower minus upper ordering.
The clock recurrence gives
\begin{dmath}\label{eq:ln-pantograph}
 {(F_{a,b}^{\AUF})'(\zeta)
 =-\ee^{b+a/2}F_{a,b}^{\AUF}(\ee^a\zeta),\qquad
 J_{a,b}'(z)=\ee^{b+a/2}J_{a,b}(\ee^a z)}
\end{dmath}.
These relations follow by differentiation and contour translation.
The alternative moment completions satisfy the same equation
for $F$, so it does not alone enforce uniqueness.

\begin{proposition}[Carrier and least-term scales]
\label{thm:ln-carrier-saddle}
For fixed $a,b$, let $L=\ln(1/r)$.
As $r\to0$, uniformly on every bounded interval of the lifted
carrier angle $\theta$,
\begin{dmath}\label{eq:ln-logarithmic-scale}
 \boxed{
 \ln|\widetilde J_{a,b}(r,\theta)|
 =-\frac{L^2}{2a}-\frac{L\ln L}{a}+O(L)
 }
\end{dmath}.
Moreover, the carrier is positive for sufficiently small $r$ and
$J_{a,b}(r)\asymp h_{a,b}(r)$. Here
$h_{a,b}(r)=\inf_{n\geq0}M_n^{(a,b)}r^n$ denotes the least-term size.
For sufficiently small $|\zeta|$ with $\RE\zeta\geq0$, optimal
truncation has remainder at most $h_{a,b}(|\zeta|)$.
\end{proposition}

The saddle estimate and proof are in \cref{app:non-gevrey-estimates}.
This carrier is smaller than every power of $r$, but larger than
$\exp(-A/r^k)$ for every fixed $A,k>0$.
Its scale reflects quadratic growth in perturbative order and is
comparable to the least term.  The result supplies no matching lower
bound for the truncation error.

Repeated monodromy does not terminate.
For $m\geq0$, define
$\widetilde J_{a,b}^{[m]}=\mathfrak M^m\widetilde J_{a,b}$.
Their inverse-Mellin clocks are
$(\ee^{-2\pi\ii t}-1)^mc_{a,b}(t)$, which retain Gaussian decay.
These lifted carriers are flat on bounded angular intervals and
linearly independent, with
$\mathfrak M^{m+1}\widetilde F_{a,b}^{\AUF}(\zeta)
=2\pi\ii\widetilde J_{a,b}^{[m]}(-\zeta)$
using the carrier logarithm in \cref{eq:ln-general-monodromy}.
Their principal planar restrictions are denoted $J_{a,b}^{[m]}$.
For fixed $m$, finite differences of carrier sheets bound their
magnitude by the same saddle scale, although cancellations can
lower it.
This hierarchy is not a sequence of successively smaller
finite-action exponential sectors.

\subsection{A physical realization in distributed relaxation}

Lognormal relaxation-time distributions occur in models of anelastic
relaxation, including Zener relaxation in alloys
\cite{NowickBerry1962}.
An ensemble of independent relaxators gives a direct realization of
the present series.  Let $u$ be elapsed time, $g\geq0$ the rate amplitude,
and $\kappa=\kappa_0\ee^{\sqrt a Y}$ with $Y\sim N(0,1)$.
The local response satisfies
\begin{dmath}\label{eq:ln-kinetic-equation}
 {\partial_uS_\kappa(g,u)=-g\kappa S_\kappa(g,u),\qquad
 S_\kappa(g,0)=1}
\end{dmath}.
Perturbing in $g$ and averaging gives
\begin{dmath}\label{eq:ln-physical-series}
 {\widehat{\langle S\rangle}(g,u)
 =\sum_{n=0}^\infty\frac{(-g\kappa_0u)^n}{n!}\ee^{a n^2/2},
 \qquad\langle S(g,u)\rangle=F_{a,0}^{\AUF}(g\kappa_0u)}
\end{dmath}.
The benchmark is therefore a short-exposure or weak-rate expansion
of a well-defined disorder-averaged response.
The Gaussian distribution of $\ln(\kappa/\kappa_0)$ supplies the
complex-order moments and hence the clock directly.
For Gaussian activation energies $E\sim N(E_0,\sigma_E^2)$, an
Arrhenius rate gives $a=\sigma_E^2/(k_BT)^2$ and
$\kappa_0=\kappa_{\rm att}\ee^{-E_0/(k_BT)}$.
The unbounded rate distribution is part of this model; a finite upper
cutoff changes the large-order class.

\section{Discussion: what is replaced, what is not}
\label{sec:comparison}
\label{sec:scope}

The central result is a direct construction of the perturbative
completion and its encoded monodromy from the same analytic clock.
In this sense AUF recovers nonperturbative information ``in one shot'':
the completion is constructed before that information is separated
into actions and fluctuation series.  The benchmarks show how such a
separation can then be performed.  Constructing an admissible clock,
and resolving all directional Stokes data, remain distinct analytic
tasks.

\subsection{What the clock constructs and what selects it}

For a W-admissible Gevrey-$s$ clock, the result is
\begin{dmath}\label{eq:comparison-summation-route}
 \widehat F\longrightarrow {c_F\longrightarrow
 \widetilde F^{\AUF}=
 \mathcal S_{k,d}\widehat F,
 \qquad k=\frac1s
 }
\end{dmath}.
The equality holds in the directions and common domains specified
by \cref{thm:mb-k-sum}.  The Barnes integral constructs the function;
the strong Gevrey estimates and Watson uniqueness identify it with
the canonical $k$-sum \cite{Balser1994}.  Thus Borel-plane continuation
is not required for this construction, although the comparison in
\cref{sec:auf-borel-laplace} recovers the usual Borel--Laplace
representation where its hypotheses hold.

This result depends on admissibility, not merely on interpolating
the coefficients.  Within the Watson class, with the same gauge,
logarithmic lift and a common admissible contour, the clocks and
their completions coincide as stated in
\cref{cor:auf-representative-independence}.  Arbitrary interpolations
do not share this uniqueness.  The analytic growth conditions are
therefore part of the construction, not optional restrictions
imposed after choosing a sum.

The non-Gevrey benchmark shows that agreement with Borel--Laplace
summation does not exhaust AUF.  Gaussian decay in imaginary order
gives a convergent completion of the lognormal series, although
every finite-order Borel transform has zero convergence radius.
Finite contour displacement establishes ordinary asymptoticity
and a flat carrier without a Watson estimate.

Here selection requires separate assumptions: the clock recurrence
and the growth restriction on its periodic interpolation factor.
Distinct positive lognormal distributions can have the same integer
moments, and the resulting completions can satisfy the same
functional equation.  Neither the coefficients nor that equation
alone select the completion.  The benchmark establishes uniqueness
within its stated clock class, rather than a universal uniqueness
claim for non-Gevrey series.

Its carrier decays faster than every power, but more slowly than
$\exp(-A/r^k)$ for every fixed $A,k>0$.  This is nonperturbative
information outside the finite-Gevrey setting used by ordinary
Borel--Laplace summation.  Summation theories adapted to
$q$-Gevrey growth remain relevant \cite{LastraSanzSendra2022}.
The distributed-relaxation realization also gives this growth
class a physical meaning: averaging a response over lognormal
rates produces precisely these perturbative coefficients.

\subsection{What the carrier contains and what remains to be resolved}

With the clockwise covering convention of
\cref{sec:auf-monodromy}, the exact relation is
\begin{dmath*}
 \mathfrak M_F(\zeta)=2\pi\ii\widetilde J_F(\eta\zeta)
\end{dmath*},
with the lift specified in \cref{eq:carrier-logarithmic-lift}.
The carrier is also the rational-gauge completion of the null clock.
In the Watson setting, $\delta>\pi$ supplies ordinary convergence and
flatness on the adjacent-sheet overlap, alongside the summation
condition $2\delta>\pi s$.
Gaussian decay supplies convergence and flatness directly in the
lognormal example.

AUF constructs this exact analytic nonperturbative variation before
any transseries decomposition.
Actions and fluctuation series are subsequently extracted from
functions already available in analytic form.
The resurgent benchmarks identify the resulting contributions
with their expected resummed expressions.

The carrier aggregates the specified one-turn variation, which is
not generally a single directional Stokes jump.
Within the ordinary inverse-Mellin class, Fourier uniqueness shows
that the principal carrier determines the selected clock.
Resolving directional Stokes relations requires comparing the
appropriate canonical sectorial sums on a common domain.

The half-line clocks in the sourced quartic model separate sourced contributions; the Gamma
carrier resolves an exponential tower; the $O(4)$ toy model yields an exact
M{\"o}bius transformation before its transseries expansion.
A fluctuation expansion may conceal exponentially smaller terms.
Alien calculus organizes elementary and iterated contributions;
a general replacement for that organization is beyond the results
proved here.

The selected completion fixes its encoded monodromy, while boundary
conditions, integration cycles or homogeneous solutions may supply
independent data.
Euler's $C\ee^{1/g}$ on sectors where $\RE(1/g)<0$ illustrates
this freedom.
These parameters belong to the global boundary-value problem,
separately from the canonical sum selected by the analytic conditions.

\subsection{Further analytic and numerical questions}

The main outstanding task is to turn the sufficient mechanisms
used here into practical criteria for constructing admissible
clocks and continuing their completions along prescribed paths.
An iterable Stokes theory also requires control of the clocks
or analytic functions generated by successive variations.
These questions concern the reach of the construction beyond
the explicit benchmarks.

The physical two-dimensional $O(4)$ sigma model is a natural
testbed, given its detailed resurgent analysis
\cite{AbbottEtAl2021,BajnokBalogVona2022,
BajnokBalogVona2023Full,BajnokBalogVona2025Complete}.
Connecting those results to a complete admissible clock and the
ordinary monodromy-carrier theorem remains an open problem,
distinct from the solvable toy model treated here.

Preliminary numerical attempts at AUF completion from partial
perturbative data have yielded promising results.
The practical task is to reconstruct a suitable clock from
finitely many coefficients; once it is available, the Barnes
integral supplies a direct quadrature scheme.
This reconstruction can be ill conditioned in some cases,
especially when small monodromy variations are sought.
Recurrences and growth estimates can constrain the inference,
although admissibility alone does not guarantee stable recovery
from partial data.
A systematic comparison with Borel--Pad{\'e} or conformal-Borel
methods should use the same coefficient count and arithmetic
precision, and test both the completed function and its
nonperturbative variations.

\section{Conclusions}
\label{sec:conclusions}

AUF provides a route to \emph{resurgence without the Borel plane}.
A W-admissible coefficient clock constructs the canonical sectorial
sum; when $\delta>\pi$, it also yields the exact one-turn monodromy
and an ordinary flat carrier on the specified overlap.
The completion and its aggregate analytic variation are obtained
``in one shot'', before actions, fluctuation series or transseries
sectors are resolved.

The lognormal example extends the construction beyond every finite
Gevrey order.
Gaussian decay in imaginary order gives a convergent completion
and a nonzero ordinary flat carrier, although every finite-order
Borel transform has zero convergence radius.
Selection rests on explicit clock conditions.
Distributed relaxation supplies a physical realization of this
coefficient growth.

The resurgent benchmarks suggest a broader correspondence between
analytic carriers and resummed transseries contributions to the
specified monodromy variation.
Establishing such a correspondence under additional resurgence
and summability hypotheses remains open.
Practical criteria for constructing clocks and controlling
directional variations are further priorities.
The physical two-dimensional $O(4)$ sigma model, with its detailed
resurgent analyses
\cite{AbbottEtAl2021,BajnokBalogVona2022,
BajnokBalogVona2023Full,BajnokBalogVona2025Complete},
is the next natural testbed for exploring these open problems.

\appendix

\section{Euler: growth estimate, null clock and continuation}
\label{app:euler-estimates}

This appendix verifies the analytic claims used in the opening
Euler example.

\subsection{The sharp Watson parameter}

Fix $0<\beta<\pi/2$.  Rotating the integration ray in
the Gamma integral through
$\operatorname{sgn}(\tau)\beta$ gives
\begin{dmath}\label{eq:euler-watson-bound}
  |\Gamma(1+\sigma+\ii\tau)|
  \leq
  \Gamma(1+\sigma)
  (\cos\beta)^{-1-\sigma}
  \ee^{-\beta|\tau|},
  \qquad \sigma>-1
\end{dmath}.
This is the Watson estimate of Gevrey order one with
$\delta=\pi+\beta$.

Conversely, fixed-line Stirling asymptotics give
\begin{dmath*}
  |\Gamma(1+\sigma+\ii\tau)|
  \sim
  \sqrt{2\pi}\,
  |\tau|^{\sigma+1/2}
  \ee^{-\pi|\tau|/2},
  \qquad |\tau|\to\infty
\end{dmath*}.
No Watson estimate with $\delta>3\pi/2$ is possible.  Hence
\begin{dmath}\label{eq:euler-sharp-W}
  \delta_\ast=\frac{3\pi}{2}
\end{dmath}.
The estimates with smaller parameters establish strong
Gevrey-one asymptotics on every proper subsector of
$|\arg g|<3\pi/2$ and carrier flatness on every proper
subsector of $|\arg z|<\pi/2$.

\subsection{The null-clock realization}

Euler's reflection formula gives
\begin{dmath}\label{eq:euler-null-clock}
  c_{J_E}(t)
  ={
  -\frac{\sin(\pi t)}{\pi}\Gamma(1+t)
  =
  \frac1{\Gamma(-t)}
  }
\end{dmath}.
Its samples vanish at every non-negative integer.
The lifted rational pairing cancels the sine factor and gives
$\widetilde J_E$; choosing $\Log z$ gives its principal planar
determination $J_E$.

The inherited estimates give carrier sectors
$|\arg z|<\beta$ for every $\beta<\pi/2$.
Their apertures approach $\pi$, but never exceed the strict
Watson uniqueness threshold for Gevrey order one.
The nonzero flat completion of the zero samples therefore does
not contradict the canonical-summation theorem.

\subsection{The continuation surface}

The differential equation has holomorphic coefficients away
from $g=0$, so the canonical germ continues along every path
in $\mathbb C^\ast$.
Let $M_E$ denote one counterclockwise turn.  Reversing the
clockwise variation gives
\begin{dmath}\label{eq:euler-monodromy}
  (M_E-I)E=2\pi\ii\ee^{1/g}
\end{dmath}.
Since the homogeneous function is single-valued,
\begin{dmath}\label{eq:euler-monodromy-n}
  M_E^nE-E
  ={
  2\pi\ii n\,\ee^{1/g},
  \qquad n\in\mathbb Z
  }
\end{dmath}.
No nonzero winding number returns the initial germ.
The maximal continuation surface over the punctured plane is
therefore
\begin{dmath}\label{eq:euler-maximal-surface}
  X_E\simeq\widetilde{\mathbb C^\ast}
\end{dmath}.

\section{Canonical AUF summation: estimates and contour transport}
\label{app:canonical-auf-proof}

\subsection{The remainder estimate and uniqueness}

Let $I_a$ denote the rational AUF pairing in
\cref{eq:framework-rational-pairing}, with the Barnes line replaced
by $a+\ii\mathbb R$ and the kernel evaluated using
$\log(-\eta\zeta)=\log\zeta+\ii\chi_\eta$.
The logarithmic coordinate is held fixed during contour displacement.

For $a=N+\tfrac12$, the Watson estimate makes the horizontal connecting
integrals vanish. The displacement rectangle is clockwise, and the
residue of the integrand $-\pi(-\eta\zeta)^t\csc(\pi t)c_F(t)$
at $t=n$ is $-\eta^n c_n\zeta^n$. Hence
\begin{dmath}\label{eq:framework-finite-residue-identity}
 \widetilde F^{\AUF}(\zeta)
 =\sum_{n=0}^{N}\eta^n c_n\zeta^n+I_{N+1/2}(\zeta)
\end{dmath}.
On a proper subsector $S'$, choose $\epsilon>0$ with
$|\vartheta+\chi_\eta|\leq\delta-\epsilon$.
Since $|\csc(\pi(N+\tfrac12+\ii\tau))|
=1/\cosh(\pi\tau)\leq2\ee^{-\pi|\tau|}$, the Watson estimate gives
\begin{dmath*}
 |I_{N+1/2}(\zeta)|
 \leq C_{S'}A^{N+1/2}
 \Gamma\bigl(1+s(N+\tfrac12)\bigr)
 (1+N)^m|\zeta|^{N+1/2}
\end{dmath*}.
The Gamma ratio relative to $\Gamma(1+sN)$ grows polynomially in $N$.
Absorb this ratio and $(1+N)^m$ into an exponential factor, bound
$|\zeta|^{1/2}$ by $r_0^{1/2}$, and separate the $n=N$ term.
The Watson bound at $t=N$ then proves
\cref{eq:framework-gevrey-remainder}, including $N=0$.

For a direction admitted by \cref{eq:framework-summable-directions},
choose $\pi s/2<\alpha<\delta-|d+\chi_\eta|$.
The sector $|\arg\zeta-d|<\alpha$ lies in the Watson sector and has
aperture greater than $\pi/k$. The strong Gevrey estimate on this
sector therefore implies $k$-summability in direction $d$, by
\cite[\S~3.1, Theorem~1, and \S~3.2]{Balser1994}.
The same theorem identifies the sum with the AUF function;
analytic uniqueness extends equality over the connected common domain.

For representative independence, use a common proper subsector of
aperture greater than $\pi s$.  The difference has zero strong
Gevrey-$s$ expansion, so Watson uniqueness makes it vanish.
For $h=c_F^{(1)}-c_F^{(2)}$, equality on the central ray
$\zeta$ with $\vartheta+\chi_\eta=0$ implies that the Fourier transform of
$\csc(\pi(\gamma+\ii\tau))h(\gamma+\ii\tau)$ vanishes.
This density is integrable by the Watson bounds.
Fourier uniqueness gives $h=0$ on the Barnes line, and the identity
theorem gives equality on its connected common analytic domain.

\subsection{Contour transport and low-order coefficients}
\label{app:auf-contour-transport}

The substitution $t=u+1$ in the lifted rational pairing,
with the logarithmic coordinate held fixed, gives
\begin{dmath}\label{eq:framework-multiplication-transport}
 \zeta\widetilde F^{\AUF}(\zeta)
 =-\frac{\eta}{2\pi\ii}
 \int_{\mathcal C_\gamma+1}
 \pi(-\eta\zeta)^t\csc(\pi t)c_F(t-1)\,\dd t
\end{dmath}.
This transports the contour without crossing a pole.
Displacing it back to $\mathcal C_\gamma$ requires analytic continuation
of the complete integrand, convergence and vanishing horizontal sides.
Surviving poles contribute residues; cancellations must be assessed
in the complete integrand.

For $\mathcal L=\sum_{j=0}^r\zeta^jP_j(\Theta_\zeta)$,
the coefficient of $\zeta^n$ in $\mathcal L\widehat F$ is
$\eta^n\sum_{j=0}^{\min(r,n)}
\eta^jP_j(n-j)c_F(n-j)$.
For $n\geq r$ this equals $\eta^n(\mathcal L_{\rm R}c_F)(n)$;
the remaining indices give separate low-order conditions.
No negative-index samples are needed.
Finally, conjugating the shift through
$\varphi_F(t)=\Gamma(1+t)c_F(t)$ gives
$\Gamma(1+t)c_F(t-1)=t\varphi_F(t-1)$,
which proves the exponential-gauge rule.

\section{Planar determinations and admissible cuts}
\label{app:auf-planar-cuts}

For a lifted cut angle $\phi$, the adapted determination is
\begin{dmath}\label{eq:planar-auf-determination}
 F_\phi^{\AUF}(\zeta)
 =-\frac1{2\pi\ii}
 \int_{\mathcal C_\gamma}
 \pi\exp[t(\log_\phi\zeta+\ii\chi_\eta)]
 \csc(\pi t)c_F(t)\,\dd t
\end{dmath}.
The integral directly supplies the part of the cut plane where
$|\arg_\phi\zeta+\chi_\eta|<\delta$.
Both banks lie in the Watson sector precisely when
$|\phi-\theta_\eta|<\delta-\pi$.
For such an admissible angle, the entire cut plane is supplied by
the integral, with locally uniform convergence up to either bank.
Other angles require separate continuation to reach any boundary
not covered by the estimate.

The logarithms on the increasing-angle banks are
\begin{dmath*}
 {
 \log_{\phi,+}(r)=\ln r+\ii(\phi-2\pi),
 \qquad
 \log_{\phi,-}(r)=\ln r+\ii\phi
 }
\end{dmath*}.
Their difference is the clockwise variation restricted to the cut.
Consequently,
\begin{dmath}\label{eq:arbitrary-cut-disc-JF}
 \Disc_\phi F_\phi^{\AUF}(r\ee^{\ii\phi})
 =2\pi\ii\widetilde J_F(r,\phi-\theta_\eta)
\end{dmath}.
The carrier logarithm is $\ln r+\ii(\phi-\theta_\eta)$.
When $|\phi-\theta_\eta|<\pi$, this lift also gives the principal
planar value $J_F(\eta r\ee^{\ii\phi})$.
In particular, $\phi=\theta_\eta$ gives
\cref{eq:natural-cut-disc-JF}, with carrier value $J_F(r)$.

Admissibility concerns the lifted angle, not only the geometric ray:
adding $2\pi$ changes the logarithmic determinations selected
by the cut.

\section{Absolute-integration bounds for the AUF--Borel comparison}
\label{app:auf-bl-integrals}

Write $M_\gamma=\int_0^\infty r^\gamma|\rho_F(r)|\,\dd r$.
On the reference determination, with
$\vartheta=\arg_{\theta_\eta}\zeta$ and
$|\vartheta+\chi_\eta|<\pi$, the absolute double integral in the
Barnes-to-density substitution is bounded by
\begin{dmath*}
 C_\gamma|\zeta|^\gamma M_\gamma
 \int_{-\infty}^{\infty}
 \ee^{-(\pi-|\vartheta+\chi_\eta|)|\tau|}\,\dd\tau
\end{dmath*}.
Fubini therefore gives \cref{eq:rational-auf-cauchy-J}.
Endpoint convergence follows because $r^\gamma$ dominates $1$ near
zero and $r^{-1}$ near infinity; the Cauchy kernel has these respective
bounds, locally uniformly away from the cut.

For the exponential density, $z=ur$ gives
$\int_0^\infty z^\gamma|\Phi_F(z)|\,\dd z
\leq\Gamma(1+\gamma)M_\gamma$ and the claimed Mellin moments.
For $\RE(\eta\zeta)<0$, absolute integration in $u,r$ is bounded by
$\int_0^\infty|\rho_F(r)|/[1-\RE(\eta\zeta)r]\,\dd r<\infty$.
The same substitution proves \cref{eq:exponential-auf-laplace}.

The exponential moment makes the density moments holomorphic for
$\RE t>\gamma$; their agreement with the clock on the original
strip extends to the integer samples.
For $|q|<\epsilon$, the summed absolute integrands are bounded by
$\ee^{|q|r}|\rho_F(r)|$, proving \cref{eq:borel-laplace-rho}.

For a direction $d$, suppose the density formula represents the
chosen Borel continuation along the whole integration ray.
Set $a_d(r;\zeta)=\RE[\ee^{\ii d}(1/\zeta-\eta r)]$.
Absolute interchange is justified if $a_d>0$ almost everywhere
on the density's support and
\begin{dmath}\label{eq:auf-bl-fubini-condition}
 \int_0^\infty\frac{|\rho_F(r)|}{a_d(r;\zeta)}\,\dd r<\infty
\end{dmath}.
This is the absolute double integral after $q=\ee^{\ii d}v$.
The inner complex integral equals $\zeta/(1-\eta\zeta r)$.
Comparison with \cref{eq:rational-auf-cauchy-J} therefore gives
$\mathcal S_d^{\rm BL}\widehat{\mathcal F}(\zeta)
=\zeta F_{\theta_\eta}^{\AUF}(\zeta)$
on the reference cut plane wherever these hypotheses hold.

For $d=-\chi_\eta$, one has $a_d=h+r$, where
$h=\RE(\ee^{\ii d}/\zeta)>0$ when $\RE(\eta\zeta)<0$.
This recovers \cref{eq:auf-bl-exact-identity} on the initial
half-plane stated in the main text.

Finally, $u=(\xi/\zeta)^k$ gives the order-$k$ monomial normalization.
On the central component, the $u$-ray lies in the right half-plane
and can be rotated to the positive axis in the Gamma integral.

\section{Non-Gevrey completion and carrier estimates}
\label{app:non-gevrey-estimates}

\subsection{Finite contour displacement}

For \cref{prop:ln-gaussian-class}, Gaussian decay dominates every
fixed lifted-angle factor, gives locally uniform convergence, and
makes the horizontal sides of finite contour displacements vanish.
Moving the rational contour to $N+1/2$ crosses residues through
index $N$ and leaves $O(|\zeta|^{N+1/2})$.
Retaining the index-$N$ term gives an $O(|\zeta|^N)$ remainder after
subtraction through index $N-1$.
The clockwise rectangle and the initial minus sign give the stated
coefficients.  Moving the pole-free carrier contour to any $d>N$
proves ordinary flatness and independence of the abscissa.
Finally,
$(\ee^{-2\pi\ii t}-1)\pi\csc(\pi t)=-2\pi\ii\ee^{-\pi\ii t}$
gives the monodromy with the assigned lift.
Fourier uniqueness on a vertical line proves that a vanishing carrier
has zero clock.

The same argument proves independence of the repeated carriers.
A vanishing finite combination would make a polynomial in
$\ee^{-2\pi\ii t}-1$ vanish identically after multiplication by
$c_{a,b}(t)$, which forces every coefficient to vanish.

\subsection{The Gaussian carrier saddle}

For \cref{thm:ln-carrier-saddle}, set
\begin{dmath}\label{eq:ln-real-saddle}
 {\Phi_L(t)=\frac a2t^2-(L-b)t-\log\Gamma(1+t),\qquad
 as-\psi(1+s)=L-b,\qquad A_s=a-\psi'(1+s)}
\end{dmath},
where $\psi=\Gamma'/\Gamma$ and $s$ is the unique large positive
solution for sufficiently large $L$.
Here $\log\Gamma(1+t)$ denotes the analytic logarithm on
$\RE t>-1$, normalized by $\log\Gamma(1)=0$.
For real $t>-1$, it equals the real logarithm
$\ln(\Gamma(1+t))$.
The logarithm of the compensated Gamma ratio below uses this
same analytic branch and vanishes at $y=0$.
On the translated contour,
\begin{dmath*}
  \widetilde J_{a,b}(r,\theta)
 =\frac{\ee^{\Phi_L(s)+\ii\theta s}}{2\pi}
 \int_{\mathbb R}\ee^{-a y^2/2-\theta y}
 \frac{\Gamma(1+s)}{\Gamma(1+s+\ii y)}
 \ee^{\ii y\psi(1+s)}\,\dd y
\end{dmath*}.
The Gamma product gives
\begin{dmath*}
 \left|\frac{\Gamma(1+s)}{\Gamma(1+s+\ii y)}\right|
 ={
 \prod_{k\geq0}\left(1+\frac{y^2}{(1+s+k)^2}\right)^{1/2}
 \leq\ee^{\psi'(1+s)y^2/2}
 }
\end{dmath*}.
The logarithm of the compensated Gamma ratio is
$\psi'(1+s)y^2/2+R_s(y)$, with
$|R_s(y)|\leq C|y|^3/s^2$ and $\RE R_s(y)\leq0$.
The polygamma series gives the first bound on the whole vertical
segment; the product inequality gives the second.
Thus $|\ee^{R_s}-1|\leq|R_s|$ and Gaussian integration yields,
uniformly for $|\theta|\leq\Theta<\infty$,
\begin{dmath}\label{eq:ln-carrier-asymptotic}
  \widetilde J_{a,b}(r,\theta)
 =\frac{\ee^{\Phi_L(s)}}{\sqrt{2\pi A_s}}
 \exp\left(\ii\theta s+\frac{\theta^2}{2A_s}\right)
 \left[1+O_{a,b,\Theta}(s^{-2})\right]
\end{dmath}.
The digamma expansion gives
$s=(L+\ln L+O_{a,b}(1))/a$ and $A_s\to a$;
substitution proves \cref{eq:ln-logarithmic-scale}.
For small $r$, $\Phi_L(s)$ is the continuous minimum of the
logarithm of $M_n^{(a,b)}r^n$.
The integer minimum is at $\lfloor s\rfloor$ or $\lceil s\rceil$,
and Taylor expansion gives
$1\leq h_{a,b}(r)/\ee^{\Phi_L(s)}\leq\ee^{a/8+O(s^{-1})}$.
This proves the least-term comparison; optimizing
\cref{eq:ln-boundary-remainder} gives the truncation bound.

\bibliographystyle{unsrtnat}
\bibliography{AUF_vs_resurgence.bib}

\end{document}